\documentclass[11pt,english,vecarrow]{extarticle}
\usepackage[T1]{fontenc}
\usepackage[utf8]{inputenc}
\usepackage[letterpaper]{geometry}
\usepackage{babel}
\usepackage{array}
\usepackage{float}
\usepackage{booktabs}
\usepackage{units}
\usepackage{amsmath}
\usepackage{amssymb}
\usepackage{graphicx}
\usepackage{setspace}
\usepackage{wasysym}

\makeatletter

\providecommand{\tabularnewline}{\\}

\numberwithin{equation}{section}
\numberwithin{figure}{section}
\numberwithin{table}{section}

\usepackage{bbm}
\usepackage{tensind}
\tensordelimiter{?}
\usepackage{slashed}
\usepackage{multicol}
\usepackage[labelfont={sc,small},textfont=small]{caption}

\makeatletter
\newcommand{\qbar}{\text{\q@bar}}
\newcommand{\q@bar}{%
  \vphantom{$\m@th q$}%
  \ooalign{%
    $\m@th q$\cr
    \hidewidth\smash{\raisebox{-1.55ex}{$\m@th\mathchar'26$}}\hidewidth\cr}%
}
\makeatother
\makeatletter
\newcommand{\lambdabar}{\text{\lambda@bar}}
\newcommand{\lambda@bar}{%
  \vphantom{$\m@th \lambda$}%
  \ooalign{%
    $\m@th \lambda$\cr
    \hidewidth\smash{\raisebox{-0.05ex}{$\m@th\mathchar'26$}}\hidewidth\cr}%
}
\makeatother

\makeatother

\begin{document}
\title{Defining geometric gauge theory to accommodate particles, continua,
and fields\\
\quad{}}
\author{\doublespacing{}\textsc{Adam Marsh}\thanks{\textsc{Electronic address: adammarsh@berkeley.edu}}}
\date{\ }
\maketitle
\begin{abstract}
Gauge theory underpins the quantum field theories of the standard
model, and in a previous paper was shown via a geometric approach
to describe classical electromagnetism in a form which approximates
QED. Here we formalize and generalize the notion of a geometric gauge
theory, then apply this framework to classical physical models, including
an improved Lagrangian for matter field electromagnetism. We find
a remarkably consistent series of actions, with straightforward limits
under which each previous one may be obtained. Ancillary benefits
include a gauge-independent Galilean Lagrangian, a geometric interpretation
for the unusual metric dependence of four-momentum, a modern treatment
of the effects of worldline variation on the four-current, a gauge
theory of gravity which includes a matter field, and consistent units
for matter field electromagnetism. 
\end{abstract}
\tableofcontents{}

\section{Introduction}

\subsection{Motivation}

Gauge theory provides a unifying geometric framework for the quantum
field theories of the standard model, but classical physical theories
are usually treated in a separate and standalone fashion. We would
like to extend the benefits of such a framework to these classical
theories, which should also provide insights into the specific limits
or approximations under which each theory may be derived from the
subsequent one. 

A previous paper \cite{Marsh-paper} introduced matter field electromagnetism,
a model for the classical theory of charged massive continua, which
was based upon a geometric view of gauge theory, and which was shown
to approximate quantum electrodynamics (QED) in the limit of a certain
quantum state. An improved Lagrangian for this theory serves as an
end point in formulating a generalized definition of geometric gauge
theory which can be extended to other classical physical theories.

A more specific motivation for such a program may be provided by \cite{Dirac},
Dirac's short book ``General Theory of Relativity,'' which despite
its title is roughly evenly split between physics and mathematics,
and which spends nearly half of its physics content on a systematic
variational presentation of relativistic continua, the electromagnetic
field, and then charged massive relativistic continua, all in the
context of gravitation as realized by curved spacetime. Dirac's approach,
however, is neither geometric nor expressed in modern mathematical
language; a substantial portion of this work is an attempt to follow
his program from such a viewpoint.

As a final motivation, but prior to any relevant definitions, we summarize
below the remarkable consistency of Lagrangians from electromagnetism
through particle mechanics when adopting the geometric viewpoint.

\[
\begin{aligned}\rho_{0}(\hbar\left\Vert \mathrm{D}_{\mu}\hat{\Phi}\right\Vert -m) & +R-\frac{1}{2}\varepsilon_{0}\left\Vert F\right\Vert ^{2} & \qquad\qquad & \textrm{Electromagnetic tensor \ensuremath{F}, matter field \ensuremath{\hat{\Phi}}}\\
\rho_{0}(m\left\Vert \hat{\underline{\Phi}}\right\Vert -m) & +R &  & \textrm{Spacetime curvature \ensuremath{R}, matter field \ensuremath{\hat{\underline{\Phi}}}}\\
\frac{1}{2}\rho_{0}(-m\left\Vert \mathrm{D}_{\tau}\sigma\right\Vert -m) &  &  & \textrm{Continua rest density \ensuremath{\rho_{0}}, direction \ensuremath{\mathrm{D}_{\tau}\sigma} }\\
\frac{1}{2}(-m\left\Vert \mathrm{D}_{\tau}\sigma\right\Vert -m) &  &  & \textrm{Minkowski particle position \ensuremath{\sigma}}\\
(\frac{1}{2}m\left\Vert \mathrm{D}_{t}\sigma\right\Vert ^{2}-m) &  &  & \textrm{Galilean particle position \ensuremath{\sigma}}
\end{aligned}
\]

\subsection{Foundations}

In formulating a generalized geometric framework for gauge theory,
the question arises as to which mathematical attributes on a manifold
are to be treated as foundational. In answering this question, we
might consider the origin of the concept of a gauge transformation,
which was introduced by Weyl \cite{Weyl 1918} in an attempt to unify
general relativity with electromagnetism. Weyl noted that the pseudo-Riemannian
geometry of general relativity is based upon a parallel transport
of tangent vectors which may ``rotate'' (Lorentz-transform) such
vectors in a path-dependent way, but which preserves their length
(corresponds to a metric compatible connection). He proposed that
the parallel transport be altered to include a path-dependent change
in vector length, and that just as the equations of general relativity
are invariant under an arbitrary change of basis in each tangent space
(e.g. a change of coordinate frame due to a change of coordinates),
the equations of physics should also be invariant under an arbitrary
change of scale in each tangent space (a conformal factor applied
to the metric), which in a later postscript he called ``gauge-invariance.''

Implicit in Weyl's approach is the idea that parallel transport is
an additional structure imposed upon a manifold which already includes
a metric; but this ignores the fact that these two attributes of a
manifold are strongly interdependent. Specifically, a given metric
implies a unique torsionless parallel transport, while a given parallel
transport with holonomy group equal to some pseudo-orthogonal group
$SO(r,s)$, absent any special symmetries, implies a unique metric
of signature $(r,s)$ with which it is compatible (up to scaling factors,
i.e. a choice of units; see \cite{Schmidt}). Moreover, the idea
of a vector being rotated after being parallel transported around
a loop is based on physical experience, e.g. the pushing of a pencil
along a globe whose projection remains tangent to a closed path; in
contrast, the idea of the pencil changing its length after such transport
is contrary to physical experience.

With these observations in mind, we choose parallel transport as our
foundational attribute. Parallel transport is arguably a more fundamental
concept than the lengths and angles specified by a metric, especially
in the context of pseudo-Riemannian metrics under which lengths may
be zero or negative. By considering parallel transport to be foundational,
we also eliminate the torsionless condition on the spacetime connection,
which acts as an obstruction to viewing it in the gauge theory context;
and (as we will see) we allow our framework to accommodate Galilean
spacetime via the lack of metric determination in the presence of
a trivial holonomy group. Finally, the notion of parallel transport
generalizes to the Ehresmann connection on fiber bundles, which defines
which points on adjacent fibers are ``the same,'' in analogy to
the way a connection defines which tangent vectors at adjacent points
are ``the same,'' and the way a matter field connection or gauge
potential defines which matter field vectors in adjacent fibers are
``the same'' in a gauge theory. 

\subsection{Overview}

In Section \ref{sec:Geometric-gauge-theory} we formalize the notion
of a geometric gauge theory in terms of smooth fiber bundles and parallel
transport, taking pains to construct a consistent framework described
in reasonably precise mathematical detail. We then define two classes
of geometric gauge theories, worldline and spacetime gauge theories,
which will accommodate the physical models of classical particles
and continua we consider. We also introduce an optional additional
geometric structure by defining embedded geometric gauge theories,
which will prove useful. 

In the subsequent sections, we apply this framework to a sequence
of physical theories, from particle mechanics to electromagnetism,
and show that we obtain the usual equations of motion (EOM), while
also making clear under what approximation or limit each theory yields
the previous one. Our geometric approach yields several ancillary
benefits, including a reference frame independent Lagrangian for non-relativistic
particles in Section \ref{sec:Particle-mechanics}, a geometric interpretation
for the unusual metric dependence of four-momentum in Section \ref{sec:Relativistic-particles},
a modern treatment of the effects of worldline variation on the four-current
in Section \ref{sec:Relativistic-continua}, a gauge theory of gravity
which includes a matter field in Section \ref{sec:General-relativity},
and an improved matter field electromagnetism with consistent units
in Section \ref{sec:Matter-field-electromagnetism}. Section \ref{sec:Summary}
then summarizes these theories in reverse, including the limits under
which each theory is obtained from the previous.

Throughout the paper we will use geometric units (as opposed to geometrized
units, see Section \ref{subsec:Units}), and detail their interpretation
and conversion to dimensionful units. We also will use the mostly
pluses spacetime metric signature, where in an orthonormal frame the
metric is $g_{\mu\nu}=\mathrm{diag}\left(-1,1,1,1\right)$, and we
will strive to standardize our index notation according to the following:
\begin{itemize}
\item Spacetime indices: $\mu,\nu,\kappa,\rho,\sigma$ 
\item Space indices: $\mathsf{i},\mathsf{j},\mathsf{k},\mathsf{m},\mathsf{n}$ 
\item Internal space indices: $a,b,c,d$ 
\item Enumeration or summation indices: $\alpha,\beta,\varsigma$
\end{itemize}
We also adopt the notation from \cite{Marsh-book} in which an arrow
decoration, e.g. $\vec{\Phi}$, indicates a vector- or $\mathbb{K}^{n}$-valued
form, where $\mathbb{K}$ is either $\mathbb{R}$ or $\mathbb{C}$,
while a check decoration, e.g. $\check{\Gamma}$ indicates an algebra-
or matrix-valued form. Finally, since we will introduce a number of
terms to categorize and keep track of our theories, we will bold these
terms when first defined.

\section{\label{sec:Geometric-gauge-theory}Geometric gauge theory}

\subsection{\label{subsec:Geometry}Geometry}

In a very rough sense, a geometric gauge theory is defined to be a
bundle of spaces with a definition of which objects on the spaces
are ``the same.'' More precisely, the core geometrical building
blocks with which we will build our theories are manifolds and the
concept of parallel transport, applied both on and between these manifolds.
Specifically, we define a \textbf{geometric gauge theory} in the most
general case to consist of a smooth fiber bundle we denote $(E_{B}^{F},B,\pi,F)$
which includes parallel transports: parallel transport of tangent
vectors on the base manifold $B$, parallel transport of tangent vectors
on each fiber $\pi^{-1}(p)\cong F$ for $p\in B$, and an Ehresmann
connection $\vec{\Gamma}$, a vector-valued 1-form on the entire bundle
$E_{B}^{F}$ which defines the vertical component of its argument,
and thus defines the parallel transport of points and tangent vectors
between adjacent fibers (see e.g. \cite{Marsh-book} pp. 247-249). 

A specific instance of a geometric gauge theory comprises a specification
of the dimensions of the base and fiber manifolds $B$ and $F$, along
with either their parallel transports or the holonomy groups of their
parallel transports. If only the holonomy group of a manifold is specified,
then that parallel transport is part of the state, and its holonomy
group is assumed to be equal to some pseudo-orthogonal group $O(r,s)$;
here $(r,s)$ is the signature, with $r$ the number of positive magnitude
vectors in an orthonormal frame and $s$ the number of negative ones,
and where $r+s$ equals the manifold dimension and thus the parallel
transport lacks any special symmetries. This means that the parallel
transport associated with a specific state is also associated with
a unique metric (up to scaling factors, i.e. a choice of units).

We will also encounter parallel transports which are specified to
be flat (trivial holonomy group). As we will see, it turns out that
this case always arises as an approximation to a curved manifold with
holonomy group of a certain signature; we therefore specify the metric
signature in this case (which is assumed to be a limit of the original
one, and would otherwise be arbitrary). 

The local trivialization maps $f\colon\pi^{-1}(p)\to F$ are assumed
to be an isomorphism with respect to the fiber metric, i.e. the fibers
are assumed to be isometric to each other and to $F$. The Ehresmann
connection is also assumed to respect the fiber metric, i.e. the diffeomorphism
between fibers it induces along a path in $B$ is required to be an
isometry. Note that this means that in the case of a general non-flat
fiber metric, the Ehresmann connection must be flat, since in general
there is only one isometry between the fiber manifolds.

\subsection{\label{subsec:States}States}

In addition to possibly including the parallel transport of $B$ or
$F$, the state of a specific geometric gauge theory comprises one
or more of the following: a bundle section 
\begin{equation}
\begin{aligned}\sigma:B & \to E_{B}^{F}\\
p & \mapsto\left.\sigma\right|_{p}\in\pi^{-1}(p),
\end{aligned}
\end{equation}
and a vertical tangent field on this section
\begin{equation}
\begin{aligned}\vec{\Phi}:B & \to VE_{B}^{F}\\
p & \mapsto\left.\vec{\Phi}\right|_{p}\in V_{\left.\sigma\right|_{p}}\equiv T_{\left.\sigma\right|_{p}}\left(\pi^{-1}(p)\right),
\end{aligned}
\end{equation}
i.e. for each point in the base, a point in the fiber over it along
with a tangent to the fiber at that point. 

We may instead consider a vertical 1-form (covector) $\underline{\Phi}$
on the section, in which case using the fiber metric we may write
$\vec{\Phi}\equiv\underline{\Phi}^{\sharp}$; as we will see, considering
the intrinsic quantity to be a 1-form can have important consequences
when varying the parallel transport. We may also consider vectors
or covectors of unit magnitude using the fiber metric, which we denote
$\hat{\Phi}$ or $\underline{\hat{\Phi}}$ and which satisfy $\left\langle \hat{\Phi},\hat{\Phi}\right\rangle =\pm1$;
in this case, their variations will be infinitesimal rotations, e.g.
the infinitesimal variation $\delta\hat{\Phi}$ will be orthogonal
to $\hat{\Phi}$.

We will use the term \textbf{matter section} to refer to the bundle
section, while the more standard term \textbf{matter field} will
refer to the vertical tangent vector field (or covector field) on
this section. The parallel transports on the base space and fibers
define the covariant derivatives of tangent vectors on each, e.g.
for $v,w\in B$
\begin{equation}
\mathrm{D}_{v}w\equiv\underset{\varepsilon\rightarrow0}{\textrm{lim}}\frac{1}{\varepsilon}\left(w\left|_{p+\varepsilon v}\right.-\parallel_{\varepsilon v}w\left|_{p}\right.\right).
\end{equation}
The Ehresmann connection defines parallel transport of points and
tangent vectors between adjacent fibers, which we explore in the next
section. 
\begin{figure}[H]
\noindent \begin{centering}
\includegraphics{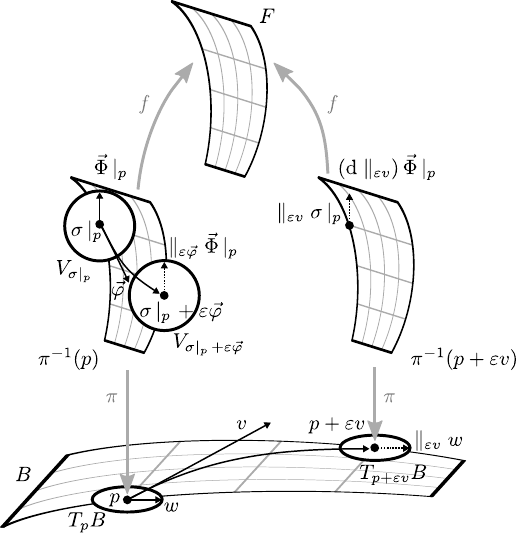}
\par\end{centering}
\caption{A geometric gauge theory is a smooth bundle with parallel transports,
with both the local trivialization maps $f$ and the diffeomorphisms
between fibers induced by the Ehresmann connection assumed to be isometries
with respect to the resulting metrics. A matter section $\sigma$
is a point on the fiber $\pi^{-1}(p)$ over each base point $p\in B$,
and a matter field $\vec{\Phi}$ is a tangent vector to the fiber
at that point; vector spaces such as the vertical tangent space $V_{\left.\sigma\right|_{p}}$
are depicted as circles. The parallel transports on the manifolds
are shown being applied to a base space tangent vector $w$ and the
fiber (vertical) tangent vector $\vec{\Phi}$ in the directions $v$
and $\vec{\varphi}$; the associated covariant derivatives are then
the difference between each vector field and its parallel transport.
The Ehresmann connection defines parallel transport of the fiber point
$\sigma$ and fiber tangent vector $\vec{\Phi}$ between fibers infinitesimally
separated by the base space vector $v$.}
\end{figure}

Note that the fibers and Ehresmann connection defined here should
not be confused with those of the principle fiber bundles used in
describing traditional gauge theories; the latter are frame bundles
which are used to describe vector bundle sections in terms of components.
Nor should our approach be confused with Kaluza-Klein theory, in which
these principal fiber bundles are taken to be additional dimensions
extending a base space which represents spacetime. The fibers in our
bundle are not abstract groups associated to a vector bundle, they
are simply manifolds associated with each point on our base manifold.

\subsection{\label{subsec:Gauge-covariant-derivatives}Gauge covariant derivatives}

We would like to define the gauge covariant derivatives of matter
sections and matter fields using the parallel transports we have available.
The Ehresmann connection $\vec{\Gamma}$ provides a (base) path-dependent
parallel transport of points across fibers; in particular, the parallel
transport of $\sigma\left|_{p}\right.$ along the path $C\in B$ from
$p$ to $q$ is defined by the horizontal lift $C_{\sigma}$, where
(see e.g. \cite{Marsh-book} Section 10.4)
\begin{equation}
\begin{aligned}C_{\sigma}\left|_{p}\right. & =\sigma\left|_{p}\right.,\\
\pi\left(C_{\sigma}\right) & =C,\\
\vec{\Gamma}\left(\dot{C}_{\sigma}\right) & =0,\\
\parallel_{C}\sigma\left|_{p}\right. & \equiv C_{\sigma}\left|_{q}\right.\in\pi^{-1}(q),
\end{aligned}
\end{equation}
and $\dot{C}_{\sigma}$ is any tangent to the curve $C_{\sigma}$.
The gauge covariant derivative of the matter section in the direction
$v=\dot{C}\left|_{p}\right.\in T_{p}B$ is then 

\begin{equation}
\begin{aligned}\mathrm{D}_{v}\sigma & \equiv\underset{\varepsilon\rightarrow0}{\textrm{lim}}\frac{1}{\varepsilon}\left(\sigma\left|_{p+\varepsilon v}\right.-\parallel_{\varepsilon v}\sigma\left|_{p}\right.\right)\\
\Rightarrow\mathrm{D}\sigma:TB & \to VE_{B}^{F}\\
v & \mapsto\mathrm{D}_{v}\sigma\in V_{\left.\sigma\right|_{p}},
\end{aligned}
\end{equation}
and $\mathrm{D}_{v}\sigma$ can be described as ``the difference
between the matter section and its parallel transport in the direction
$v$.'' Note that for a fixed vector field $v$ on $B$, $\mathrm{D}_{v}\sigma$
is itself a matter field.

Applied to the entire fiber, the parallel transport defined by the
Ehresmann connection provides a (base) path-dependent diffeomorphism
between fibers, which is furthermore assumed to be an isometry with
respect to the fiber metric. The differential of this isometry then
provides a path-dependent parallel transport of vertical tangents
across fibers which preserves length and angles; in particular, for
the matter field we have
\begin{equation}
\begin{aligned}\left(\mathrm{d}\parallel_{C}\right)\vec{\Phi}\left|_{p}\right. & \in V_{\parallel_{C}\sigma\left|_{p}\right.}.\end{aligned}
\end{equation}
For infinitesimal curves, we may then use the parallel transport on
the fiber to yield a vector which may be compared to the matter field
\begin{equation}
\begin{aligned}\parallel_{\varepsilon v}\vec{\Phi}\left|_{p}\right.\equiv\parallel_{\varepsilon D_{v}\sigma}\left(\left(\mathrm{d}\parallel_{\varepsilon v}\right)\vec{\Phi}\left|_{p}\right.\right) & \in V_{\sigma\left|_{p+\varepsilon v}\right.},\end{aligned}
\end{equation}
which defines the gauge covariant derivative of the matter field

\begin{equation}
\begin{aligned}\mathrm{D}_{v}\vec{\Phi} & \equiv\underset{\varepsilon\rightarrow0}{\textrm{lim}}\frac{1}{\varepsilon}\left(\vec{\Phi}\left|_{p+\varepsilon v}\right.-\parallel_{\varepsilon v}\vec{\Phi}\left|_{p}\right.\right)\\
\Rightarrow\mathrm{D}\vec{\Phi}:TB & \to VE_{B}^{F}\\
v & \mapsto\mathrm{D}_{v}\vec{\Phi}\in V_{\left.\sigma\right|_{p}},
\end{aligned}
\end{equation}
and $\mathrm{D}_{v}\vec{\Phi}$ can be described as ``the difference
between the matter field and its parallel transport in the direction
$v$.'' Again, for a fixed vector field $v$ on $B$, $\mathrm{D}_{v}\vec{\Phi}$
is itself a matter field. 
\begin{figure}[H]
\noindent \begin{centering}
\includegraphics{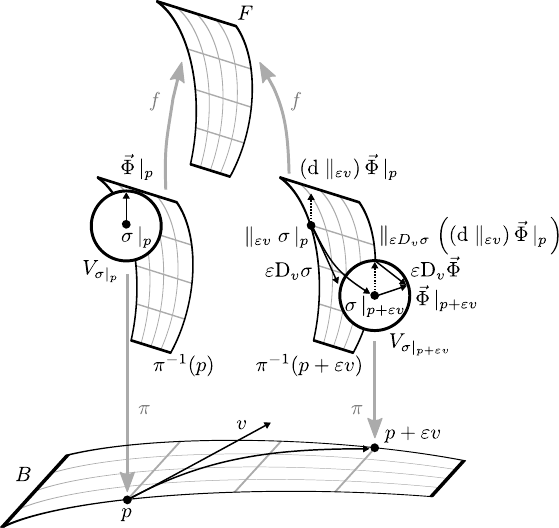}
\par\end{centering}
\caption{The gauge covariant derivatives of the section and the field in a
direction $v\in T_{p}B$ are defined by the parallel transports on
and between the fibers as depicted.}
\end{figure}

\subsection{\label{subsec:Choices-of-gauge}Choices of gauge}

Depending on the symmetries of our specific theory, we may also be
able to define a set of preferred coordinates or preferred frames
on patches of the base space or fibers which simplify computations.
We will call a smoothly defined choice of such coordinates on the
fibers a choice of \textbf{matter section gauge}; it enables us to
write each point in our section in terms of these coordinates. Similarly,
we will call a smoothly defined choice of such a frame on the fibers
a choice of \textbf{matter field gauge}; it enables us to write each
tangent vector in terms of frame components. A gauge transformation
is then a change of preferred coordinates or frame. 

In particular, if the fiber metric has signature $(r,s)$ where the
dimension of the fiber manifold is $r+s=n$, a choice of orthonormal
frame lets us write the gauge-independent vector-valued 1-form $\mathrm{D}\vec{\Phi}$
in terms of the gauge-dependent $\mathbb{R}^{n}$-valued 0-form $\vec{\Phi}$
and the \textbf{matter field connection}, a gauge-dependent matrix-valued
1-form $\check{\Gamma}_{A}$: 

\begin{equation}
\begin{aligned}\mathrm{D}\vec{\Phi} & \equiv\mathrm{d}\vec{\Phi}+\check{\Gamma}_{A}\vec{\Phi}\\
\Rightarrow\check{\Gamma}_{A} & :TB\to so(r,s).
\end{aligned}
\end{equation}
These expressions may also be written in terms of components in the
chosen frame on the fibers. 

In the following we define two classes of geometric gauge theories,
their distinction being whether the base space represents a worldline
or spacetime. In the latter type, we will allow the fiber manifold
and tangent space to be complex, enabling us to define the traditional
gauge theory machinery. In particular, the matter field connection
above becomes a gauge-dependent matrix-valued 1-form on $B$ 

\begin{equation}
\begin{aligned}\check{\Gamma}_{A} & :TB\to su(n).\end{aligned}
\end{equation}

\subsection{\label{subsec:Worldline-gauge-theories}Worldline gauge theories}

The initial class of geometric gauge theories we consider will consist
of a trivial bundle $(E_{\Lambda}^{M},\Lambda,\pi,M)$ over a one
dimensional base space, which with a choice of origin, units, and
positive direction we can assume is $\Lambda\cong\mathbb{R}$, establishing
a coordinate $\lambda$ on $\Lambda$ whose coordinate frame has unit
length. This coordinate lets us label each fiber $M_{\lambda}$, and
a matter section $\sigma$ defines a path on the entire space $E_{\Lambda}^{M}=\Lambda\times M$.
We will call a member of this class of geometric gauge theories a
\textbf{worldline gauge theory}.

Since there is only one possible path between any two points on the
base space, the parallel transport along $\Lambda$ using the Ehresmann
connection provides a unique isometry between fibers, which allows
us to project down to the fiber $E_{\Lambda}^{M}\rightarrow M$ (in
contrast to the bundle projection $E_{\Lambda}^{M}\rightarrow\Lambda$)
while preserving the fiber metric. We will refer to $M$ under this
projection as the \textbf{collapsed bundle}. A matter section $\sigma$
projects down to a curve in $M$ parametrized by the coordinate $\lambda$,
which we denote $x_{\sigma}(\lambda)$. The matter field may then
be defined to be the gauge covariant derivative
\begin{equation}
\vec{\Phi}\equiv\mathrm{D}_{\lambda}\sigma=\partial_{\lambda}x_{\sigma},
\end{equation}
the tangent vector to the parametrized curve in $M$. All tangent
vectors on $M$ have an inner product provided by the fiber metric,
which we denote $g$, and which is defined by the fiber parallel transport
(which may or may not be part of the state) and a choice of units. 

We will refer to a matter section gauge (preferred coordinates on
each fiber) that respects parallel transport (the coordinate functions
are parallel transports of each other), as a \textbf{parallel gauge};
such coordinates also project down to $M$, allowing the matter field
to be written
\begin{equation}
\Phi^{\mu}=\partial_{\lambda}x_{\sigma}^{\mu}.
\end{equation}
A gauge transformation of a parallel gauge is then an identical change
of coordinates on each fiber $M_{\lambda}$, in order to preserve
the parallel transport of coordinate functions across fibers. 

Now, in the specific case in which the fibers are flat, we have an
alternative way to define an isometry between fibers: a matter section
gauge, a smoothly defined choice of coordinates on each fiber, which
then project down to $M$. However, the associated projection of the
matter section $\sigma$ down to $M$ is no longer gauge-independent,
since a change of gauge is arbitrary. We will refer to $M$ under
this projection as the \textbf{gauge collapsed bundle}. The components
$\sigma^{\mu}$ of a matter section due to the fiber coordinates are
equal to the coordinates of the projected curve $x_{\sigma}^{\mu}(\lambda)$
in $M$. The matter field may still be defined to be the gauge covariant
derivative 
\begin{equation}
\vec{\Phi}\equiv\mathrm{D}_{\lambda}\sigma,
\end{equation}
but it is no longer equal to $\partial_{\lambda}x_{\sigma}^{\mu}$,
and like the curve $x_{\sigma}(\lambda)$, its projection down to
$M$ is not invariant under gauge transformations. We will see our
only use cases of this gauge-dependent projection in Section \ref{sec:Particle-mechanics}
on particle mechanics.
\begin{figure}[H]
\noindent \begin{centering}
\includegraphics{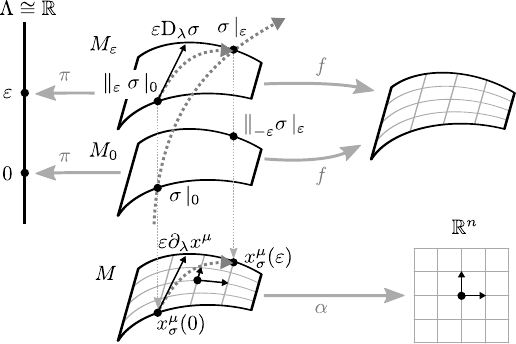}
\par\end{centering}
\caption{A worldline gauge theory is a trivial bundle over $\mathbb{R}$ with
fiber an $n$-dimensional pseudo-Riemannian manifold $M^{n}$ and
matter section $\sigma$. If we smoothly define coordinates $\alpha_{\lambda}\colon M_{\lambda}\to\mathbb{R}^{n}$
on each fiber, a matter section $\sigma$ is a path in the entire
space $\mathbb{R}\times M$ which projects down via these coordinates
to a parametrized curve $x_{\sigma}^{\mu}(\lambda)$ in the gauge
collapsed bundle $M$. In the figure, we choose a parallel gauge,
which associates points in the fibers via parallel transport and selects
coordinates $\alpha$ on the fibers which respect this association.
The covariant derivative is then $\mathrm{D}_{\lambda}\sigma=\partial_{\lambda}x_{\sigma}^{\mu}$,
the tangent vector to the parametrized curve in the collapsed bundle
$M$, which also serves as the matter field $\vec{\Phi}$.}
\end{figure}

\subsection{\label{subsec:Spacetime-gauge-theories}Spacetime gauge theories}

The second class of geometric gauge theories we consider will consist
of a bundle $(E_{M}^{X},M,\pi,X)$ over a Lorentzian base space representing
spacetime, which in this class of theories we denote $M$. The fact
that the symbol for the base space $M$ coincides with that of the
collapsed bundle of a worldline gauge theory is no accident, as we
will first see in Section \ref{sec:General-relativity} on general
relativity. For the same reason, we use Greek indices for components
of points and vectors on $M$, while introducing Latin indices for
components on the fiber, which we denote $X$ and whose parallel transport
we assume to result in a Riemannian metric. We will call a member
of this class of geometric gauge theories a \textbf{spacetime gauge
theory}. 

If the base space parallel transport is flat, we will assume it has
a Minkowski metric $g$ with the ``mostly pluses'' signature $\left(3,1\right)$,
dependent only upon a choice of units. If it is not flat, we will
assume that it has holonomy group $SO(3,1)$, resulting in a connection
in an orthonormal frame

\begin{equation}
\begin{aligned}\check{\Gamma} & :TM\to so(3,1),\end{aligned}
\end{equation}
which we call the \textbf{spacetime connection}. We will not assume
that this connection is torsion-free; up to a choice of units it determines
a unique Lorentzian metric $g$ on $M$. 

This class of geometric gauge theory will usually only have a matter
field explicitly identified, with the matter section, Ehresmann connection,
and fiber connection left unspecified. This results in an arbitrary
matter field gauge covariant derivative, and thus an arbitrary matter
field connection (as a concrete example, we may take the fiber to
be a sphere, with the matter section horizontal in directions with
no curvature, while mapping any holonomy loop to a path on the fiber
which results in the desired curvature). It also allows the bundle
to be viewed as a vector bundle $(E,M,\pi_{E},\mathbb{R}^{n})$, whose
vector space fiber over $p\in M$ is the vertical tangent space in
an orthonormal fiber frame 
\begin{equation}
\begin{aligned}\pi_{E}^{-1}(p) & =V_{\left.\sigma\right|_{p}}\\
 & \equiv V_{p},
\end{aligned}
\end{equation}
which is called the \textbf{internal space}, with the matter field
$\vec{\Phi}$ an element of this vector space. The local trivialization
maps are isomorphisms from $V_{p}$ to $\mathbb{R}^{n}$ (or $\mathbb{C}^{n}$
if $X$ is complex), and a choice of a gauge is a choice of orthonormal
frame smoothly defined on each manifold fiber, which results in a
choice of orthonormal basis in each vertical tangent space.

Now, any variation of the fiber parallel transport would have to occur
across all fibers simultaneously in order to keep them isometric,
and would not effect a general variation of the matter field parallel
transport; for example, it could not change the matter field gauge
covariant derivative in a direction in which the associated matter
section is horizontal. We therefore accomplish variation of the matter
field parallel transport via variation of the Ehresmann connection,
which as we can see from our previous example of a spherical fiber,
results in an arbitrary varied parallel transport which still preserves
the length of $\vec{\Phi}$.  

This class of geometric gauge theories thus coincides with the more
standard definition of a gauge theory, but retains the fibers to which
each vector space is tangent. Using Greek components based on the
coordinates on the base space $M$, and Latin components based on
the frame (gauge) on the fibers, we may write
\begin{align}
\begin{aligned}\mathrm{D}\vec{\Phi} & =\mathrm{d}\vec{\Phi}+\check{\Gamma}_{A}\vec{\Phi},\\
\mathrm{D}\Phi^{a} & =\partial_{\mu}\Phi^{a}+\left(\Gamma_{A}\right)^{a}{}_{b\mu}\Phi^{b},\\
\check{R}_{A} & =\mathrm{d}\check{\Gamma}_{A}+\check{\Gamma}_{A}\wedge\check{\Gamma}_{A},
\end{aligned}
\end{align}
where $\check{R}_{A}$ is the curvature of the connection $\check{\Gamma}_{A}$.
If we allow the fiber to be complex, then the gauge covariant derivative
is written in terms of the $\mathbb{C}^{n}$-valued 0-form $\vec{\Phi}$
and the hermitian matrix-valued 1-form 
\begin{equation}
\begin{aligned}\check{A} & \equiv\frac{i}{\qbar}\check{\Gamma}_{A},\\
\qbar & \equiv\frac{q}{\hbar},
\end{aligned}
\end{equation}
which is the gauge potential 
\begin{align}
\begin{aligned}\mathrm{D}\vec{\Phi} & =\mathrm{d}\vec{\Phi}-i\qbar\check{A}\vec{\Phi},\\
\mathrm{D}\Phi^{a} & =\partial_{\mu}\Phi^{a}-i\qbar A^{a}{}_{b\mu}\Phi^{b}
\end{aligned}
\end{align}
and defines the field strength
\begin{equation}
\begin{aligned}\check{F} & \equiv\frac{i}{\qbar}\check{R}_{A}\\
\Rightarrow\check{F} & =\mathrm{d}\check{A}-i\qbar\check{A}\land\check{A}.
\end{aligned}
\end{equation}
The reduced Planck constant $\hbar$ and the coupling constant $q$
will be given geometric interpretations in Section \ref{sec:Matter-field-electromagnetism}.
\begin{figure}[H]
\noindent \begin{centering}
\includegraphics{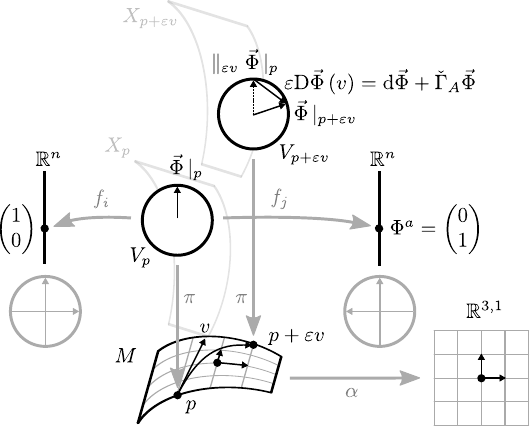}
\par\end{centering}
\caption{A spacetime gauge theory is a smooth bundle over spacetime $M$ with
manifold fiber $X$. Only a matter field $\vec{\Phi}$ (a vertical
tangent vector at a point on each fiber) is specified, leaving the
matter field connection $\check{\Gamma}_{A}$ arbitrary via specification
of the Ehresmann connection, and allowing the matter field to be viewed
as the section of a vector bundle with fiber $\mathbb{R}^{n}$, and
fiber isomorphisms $f_{i}:V_{p}\to\mathbb{R}^{n}$. If we choose a
gauge, an orthonormal frame on each $X$, we specify a basis for each
vertical tangent space $V$ which defines $f_{i}$; a matter field
$\vec{\Phi}$ can then be written in terms of components $\Phi^{a}$,
with a gauge transformation being a smooth change of orthonormal basis
for each $V$ and the gauge covariant derivative preserving matter
field length. If we allow the fiber to be complex, the vector space
fiber isomorphisms are $f_{i}:V_{p}\to\mathbb{C}^{n}$ and the matter
field connection is expressed in terms of the gauge potential $\check{A}$.}
\end{figure}

\subsection{\label{subsec:Embedded-worldline-gauge-theories}Embedded worldline
gauge theories}

In a worldline gauge theory, the matter section $\sigma$ may be viewed
as a curve $x_{\sigma}(\lambda)$ in the collapsed  bundle $M$, parametrized
by a coordinate $\lambda$ whose coordinate frame has unit length
on $\Lambda$. This curve may then be identified with the base space
$\Lambda$ itself, embedded in $M$, so that the frame on $\Lambda$
may be written in a parallel gauge (parallel coordinate functions
on the fibers) as 
\begin{equation}
\begin{aligned}\frac{\partial}{\partial\lambda}\in\Lambda & =\partial_{\lambda}x_{\sigma}\in M\\
 & =\partial_{\lambda}x_{\sigma}^{\mu}\frac{\partial}{\partial x^{\mu}}.
\end{aligned}
\end{equation}
We use this view to confer additional geometric structure on the theory,
by requiring that for some choice of metric scaling factor on $M$,
this embedding be isometric. The name we give this geometric structure
(including the requirement of a parallel gauge) is an \textbf{embedded
worldline gauge theory}.

A non-null isometric embedding of $\Lambda$ means that the metric
on $\Lambda$, and thus the parametrization $\lambda$ up to a choice
of origin, is induced by that of $M$, so that tangent vectors $\partial_{\lambda}x_{\sigma}$
have a constant unit length in $M$. In keeping with our view of parallel
transport as the fundamental quantity, we further assume that under
a uniform scaling of the metric on $M$, which leaves parallel transport
unchanged, the vector $\partial_{\lambda}x_{\sigma}$ remains constant,
therefore changing its length; in other words, the choice of units
on $M$ introduces a constant proportionality factor between the metrics
on $\Lambda$ and $M$. However, an infinitesimal variation of the
metric on $M$, which stems from a variation of parallel transport,
will in fact change $\lambda$ and thus $\partial_{\lambda}x_{\sigma}$
in order to leave its length unchanged under the isometric embedding.

Note that the embedding in an embedded worldline gauge theory is associated
with a specific matter section, and is therefore part of the state.
\begin{figure}[H]
\noindent \begin{centering}
\includegraphics{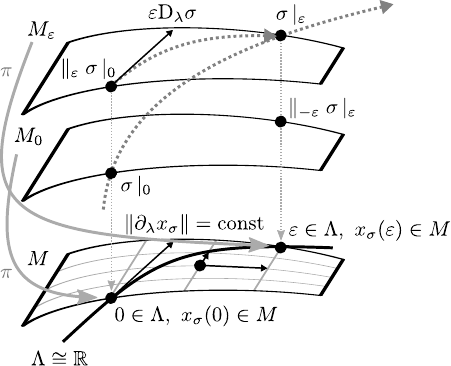}
\par\end{centering}
\caption{An embedded worldline gauge theory isometrically embeds the base space
$\Lambda$ in the collapsed  bundle $M$ under a parallel gauge, causing
the tangent vectors $\partial_{\lambda}x_{\sigma}$ to have constant
length; scaling of the metric (a choice of units) on $M$ leaves $\partial_{\lambda}x_{\sigma}$
unchanged, this changing this constant, while variation of the metric
changes $\partial_{\lambda}x_{\sigma}$ to keep its length the same.
}
\end{figure}

\subsection{\label{subsec:Embedded-spacetime-gauge-theories}Embedded spacetime
gauge theories}

Instead of embedding the base space in the collapsed bundle, we may
consider embedding the fibers in the base space; this requires that
the fiber be of smaller or equal dimension and compatible signature.
In particular, if both base space and fiber are Lorentzian manifolds
of the same dimension, then the fiber $X_{p}$ over every point $p\in M$
may be considered to be isometric to $M$ itself, with the parallel
matter section over $p$ the corresponding point in $X_{p}$, and
a matter field a vector field on $M$ whose parallel transport is
that of the spacetime connection on $M$. We will see this type of
embedded geometric gauge theory used in an alternative formulation
of general relativity in Section \ref{sec:General-relativity}.

In the case of a spacetime gauge theory, $X$ is Riemannian, and thus
if $n<4$ we may isometrically embed each $X_{p}$ as a space-like
submanifold of $M$. Just as we could not consider parallel transport
of worldline tangents to be tangent vector parallel transport in the
collapsed bundle of an embedded worldline gauge theory (since it would
not remain tangent to the worldline), we cannot consider matter field
parallel transport here to be tangent vector parallel transport in
$M$; instead the matter field gauge covariant derivative remains
based on the Ehresmann connection between fibers, and thus arbitrary,
but we have the additional structure of a space-like vector subspace
of the tangent space $V_{p}\subset T_{p}M$ at each point, in which
the matter field takes its values. The name we give this geometric
structure is an \textbf{embedded spacetime gauge theory}. 

Since the fibers must remain isometric, we assume that a variation
of the metric on $M$ alters the fibers to keep the $V_{p}$ and fiber
metrics unchanged, therefore also leaving the matter field and matter
field connection unchanged. As in an embedded worldline gauge theory,
a scaling of the metric proportionally changes the length of $\vec{\Phi}$,
but does not change the length of a matter field $\hat{\Phi}$ which
is defined to be of unit length. We also note as with embedded worldline
gauge theories, the embedding in an embedded spacetime gauge theory
is associated with a specific matter section, and therefore is part
of the state. 
\begin{figure}[H]
\noindent \begin{centering}
\includegraphics{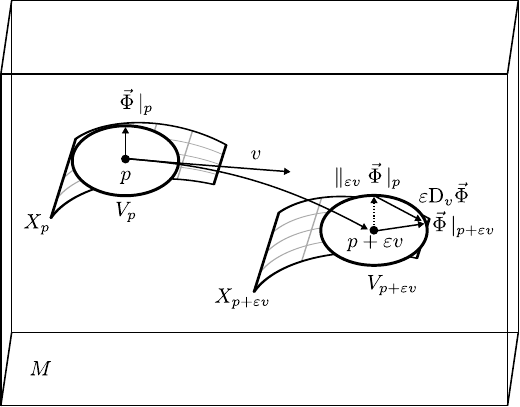}
\par\end{centering}
\caption{An embedded spacetime gauge theory with a Riemannian fiber space $X^{n}$
of lower dimension than the Lorentzian spacetime manifold $M$ remains
a vector bundle, but features the additional geometric structure of
a space-like subspace of the tangent space $V_{p}\subset T_{p}M$
at each point in which the matter field $\vec{\Phi}$ takes its values.
In the figure we have $n=2$, and note that $v\in T_{p}M$ does not
necessarily lie in $V_{p}$. }
\end{figure}

\subsection{\label{subsec:Units}Units}

With a concrete geometry associated with our physical theories, we
may express every quantity in terms of the choices of scaling factors
in the metrics we select compatible with the parallel transport defined
on our manifolds, i.e. in units of length. Expressions in terms of
these chosen scales will be called \textbf{geometric units}. Since
our geometric units are specific to geometric gauge theories, we will
detail their conversion to dimensionful units, specifically SI units.

Our geometric units should not be confused with geometrized units,
which also express quantities in terms of length, but do so by setting
certain physical constants to unity, as opposed to positing that these
lengths originate from an underlying geometric model. Geometrized
units are a system of natural units, which when extended to theories
which include electromagnetism typically fix a length unit; geometric
units do not fix a length unit, but only express quantities in terms
of one or more base physical dimensions corresponding to manifold
metrics. In all of the theories we cover beyond particle mechanics,
all geometric units will be expressed in terms of a single physical
dimension $\mathsf{L}$.

\section{\label{sec:Particle-mechanics}Particle mechanics}

In particle mechanics, the action is usually expressed in terms of
velocity, which is dependent upon the inertial reference frame. In
the geometric gauge theory formulation, this is restated as velocity
not being gauge-invariant. Here we present a gauge-invariant action
which results in Newton's laws as the matter section EOM, and also
identifies the center of mass frame as the parallel gauge via variation
of the gauge potential.

\subsection{\label{subsec:The-Galilean-bundle}The Galilean bundle}

The classical mechanics of point particles may be described by a worldline
gauge theory $(\mathcal{G},T,\pi,S)$; following e.g. Penrose \cite{Penrose}
we call this the \textbf{Galilean bundle}. The base space $T\cong\mathbb{R}$
is time, and the \textbf{space fiber} $S_{t}\cong\mathbb{R}^{3}$
over each point $t\in T$ is space at that instant in time. The coordinate
$t$ on $T$ is assumed to have a coordinate frame which has unit
length, and thus comprises a choice of origin and units of time. Matter
sections are point particles, and are $S$-valued 0-forms $\sigma$
on $T$. Viewed as a path in the entire space $\mathcal{G}$, a specific
matter section $\sigma_{1}$ is the worldline of particle 1, and points
on the worldline are events. 

Since the space fiber is flat, the metric is determined by an arbitrary
choice of pseudo inner product at any point, which applies to all
other points via the path-independent parallel transport. We make
a choice of Riemannian metric (and thus a unit of spatial length),
which then extends to a Riemannian metric across all fibers via the
isometries of the Ehresmann connection, which we view as part of the
state. 

A matter section gauge is a reference frame, and is defined to be
a smoothly defined coordinate chart $\alpha\colon S_{t}\to\mathbb{R}^{3}$
for space at each point in time whose coordinate frame is orthonormal;
it allows us to express the matter section $\sigma_{1}$ projected
down to the gauge collapsed bundle as $x_{1}^{\mathsf{i}}\left(t\right)\in S$.
The origin of each of these space coordinates defines a section of
the bundle; this section can be thought of as the worldline of an
observer particle $\sigma_{0}$ from whose reference point particle
positions are measured. In particular, the observer particle defined
by a parallel gauge, which we denote $\sigma_{\parallel}$, may be
both rotated and displaced over time in a different gauge. A change
of reference frame is then a gauge transformation on this bundle,
a transformation of the spatial coordinates at each time which keeps
the coordinate frame orthonormal. 

We will call the gauge collapsed  bundle $S$ associated with a reference
frame on the Galilean bundle \textbf{Galilean space}, and the parametrized
path in $S$ corresponding to a worldline in $\mathcal{G}$ will be
called a \textbf{trajectory}. Unlike with the worldline, the tangent
of the trajectory $\partial_{t}x_{1}^{\mathsf{i}}$ may vanish; in
particular, the observer particle has constant coordinates which are
all zero, hence its trajectory is a point at the origin. More importantly,
the trajectory does not remain constant under a change of reference
frame, and therefore unlike the worldline is not gauge-invariant,
and hence cannot be viewed as an intrinsic object. 

A change of reference frame applied to a point particle is a gauge
transformation applied to a matter section, and may be written as
a transformation
\begin{equation}
\begin{aligned}\gamma\colon\alpha\left(\sigma_{1}\right) & \to\alpha^{\prime}\left(\sigma_{1}\right)\\
x_{1}^{\mathsf{i}}\left(t\right) & \mapsto\left(x_{1}^{\mathsf{i}}\right)^{\prime}\left(t\right)
\end{aligned}
\end{equation}
to a new coordinate chart, and changes the actual trajectory on $S$.
Given a reference frame on the Galilean bundle, we may then define
the velocity and acceleration for each particle in Galilean space:

\begin{equation}
\begin{aligned}v_{1}^{\mathsf{i}} & \equiv\partial_{t}x_{1}^{\mathsf{i}},\\
a_{1}^{\mathsf{i}} & \equiv\partial_{t}^{2}x_{1}^{\mathsf{i}}.
\end{aligned}
\end{equation}
Note however that these quantities are dependent upon the reference
frame, i.e. they are not gauge-independent quantities. 

We define the more restrictive \textbf{inertial Galilean bundle},
which we denote identically and which becomes our primary model, by
choosing preferred spatial coordinates which result in inertial reference
frames. More precisely, Newton's first law posits the existence of
an inertial reference frame, a set of spatial coordinates in which
free particles have straight worldlines; we then define a change of
inertial reference frame (AKA Galilean transformation) to be a transformation
$\gamma$ to a new choice of gauge on the Galilean bundle which keeps
constant velocities constant. If we treat the values of $\alpha\colon S_{t}\to\mathbb{R}^{3}$,
which are points in the manifold $\mathbb{R}^{3}$, as vectors in
the vector space $\mathbb{R}^{3}$, then omitting reflections, $\gamma$
has a rotational component $\gamma_{\mathrm{r}}\in SO(3)$ which is
constant on $T$, and a translational (displacement) component $\gamma_{\mathrm{d}}\in\mathbb{R}^{3}$
which is linear on $T$:
\begin{equation}
\begin{aligned}\gamma_{\mathrm{d}}\left(t\right) & \equiv\gamma_{\mathrm{z}}+\gamma_{\mathrm{v}}\left(t-t_{\varnothing}\right)\\
\Rightarrow\left(x_{1}^{\mathsf{i}}\right)^{\prime}\left(t\right) & =(\gamma_{\mathrm{r}})^{\mathsf{i}}{}_{\mathsf{j}}x_{1}^{\mathsf{j}}\left(t\right)+\left(\gamma_{\mathrm{v}}\right)^{\mathsf{i}}\left(t-t_{\varnothing}\right)+\left(\gamma_{\mathrm{z}}\right)^{\mathsf{i}}.
\end{aligned}
\end{equation}
Thus 
\begin{equation}
\left(v_{1}^{\mathsf{i}}\right)^{\prime}=(\gamma_{\mathrm{r}})^{\mathsf{i}}{}_{\mathsf{j}}v_{1}^{\mathsf{j}}+\left(\gamma_{\mathrm{v}}\right)^{\mathsf{i}},
\end{equation}
which is constant if $v_{1}^{\mathsf{j}}$ is. A \textbf{Galilean
boost} is a change of inertial frame which implies a new observer
coincident at the origin and moving at velocity $\gamma_{\mathrm{v}}$
(which therefore transforms coordinates by $x^{\prime}=-\gamma_{\mathrm{v}}t$).
We call the gauge collapsed bundle of the inertial Galilean bundle
\textbf{inertial Galilean space} (AKA Newtonian space), and assume
that the parallel observer particle $\sigma_{\parallel}$ has an inertial
worldline. 
\begin{figure}[H]
\noindent \begin{centering}
\includegraphics{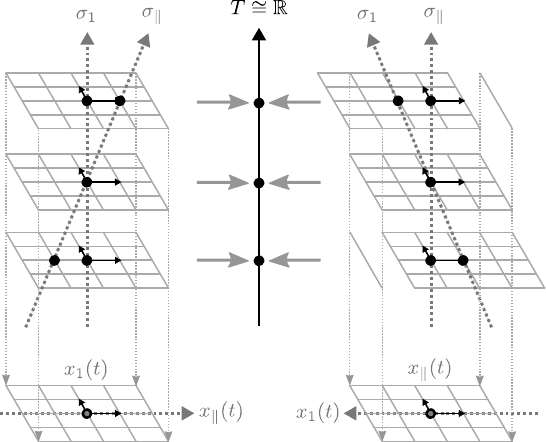}
\par\end{centering}
\caption{Inertial Galilean space is the gauge collapsed bundle of the inertial
Galilean bundle in a given inertial reference frame. The trajectories
of particles in inertial Galilean space are parametrized by absolute
time, but do not remain constant under a Galilean boost.}
\end{figure}

\subsection{\label{subsec:PM-gauge-covariant-derivative}The gauge covariant
derivative }

Since we assume that the parallel observer particle $\sigma_{\parallel}$
has an inertial worldline, the parallel transport in the inertial
reference frame of any other free particle $\sigma_{1}$ at $t=t_{0}$
is just the original coordinates plus $v_{\parallel}^{\mathsf{i}}$
times the time transported, i.e. for an infinitesimal parallel transport
we have
\begin{equation}
\parallel_{\varepsilon}\left(\left.x_{1}^{\mathsf{i}}\right|_{t_{0}}\right)=x_{1}^{\mathsf{i}}\left|_{t_{0}}\right.+\varepsilon\left.v_{\parallel}^{\mathsf{i}}\right|_{t_{0}}.
\end{equation}
Hence the gauge covariant derivative (matter field) in any inertial
frame can be written

\begin{equation}
\begin{aligned}\mathrm{D}_{t}x_{1}^{\mathsf{i}} & =\underset{\varepsilon\rightarrow0}{\textrm{lim}}\frac{1}{\varepsilon}\left(\left.x_{1}^{\mathsf{i}}\right|_{t_{0}+\varepsilon}-\parallel_{\varepsilon}\left(\left.x_{1}^{\mathsf{i}}\right|_{t_{0}}\right)\right)\\
 & =\underset{\varepsilon\rightarrow0}{\textrm{lim}}\frac{1}{\varepsilon}\left(\left.x_{1}^{\mathsf{i}}\right|_{t_{0}+\varepsilon}-\left(\left.x_{1}^{\mathsf{i}}\right|_{t_{0}}+\varepsilon\left.v_{\parallel}^{\mathsf{i}}\right|_{t_{0}}\right)\right)\\
 & =\partial_{t}x_{1}^{\mathsf{i}}-v_{\parallel}^{\mathsf{i}}\\
 & =v_{1}^{\mathsf{i}}-v_{\parallel}^{\mathsf{i}}.
\end{aligned}
\end{equation}
Since the parallel transport of the matter field value $\mathrm{D}_{t_{0}}x_{1}^{\mathsf{i}}$
simply displaces the vector to the matter section point $\left.x_{1}^{\mathsf{i}}\right|_{t_{0}+\varepsilon}$,
the gauge covariant derivative is just the time derivative; hence
taking the gauge covariant derivative twice yields 
\begin{equation}
\mathrm{D}_{t}^{2}x_{1}=a_{1},
\end{equation}
since $v_{\parallel}^{\mathsf{i}}$ is constant, so that the acceleration
of a particle is a gauge-independent quantity in our preferred inertial
coordinates. 

\subsection{\label{subsec:The-Galilean-nested-bundle}The Galilean nested bundle}

If we view the total space of the inertial Galilean bundle $\mathcal{G}$
as a manifold, then a given inertial reference frame is a diffeomorphism
$\mathcal{G}\to T\times\mathbb{R}^{3}$, and the metric on $\mathcal{G}$
induced by those on $T$ and each fiber $\mathbb{R}^{3}$ make it
isometric to $\mathbb{R}^{4}$. We will call the manifold $\mathcal{G}$
equipped with this metric the \textbf{Galilean spacetime manifold}.
Again treating the points in the manifold $\mathbb{R}^{4}$ as vectors
and omitting reflections, we can view inertial Galilean transformations
as acting on the entire inertial Galilean bundle, taking values in
the special Galilean group 
\begin{equation}
\mathrm{SGal}(3)\equiv\mathbb{R}^{4}\rtimes\mathbb{R}^{3}\rtimes SO(3),
\end{equation}
which is a semidirect product whose factor groups are global displacement,
relative frame velocity (boost), and spatial rotation.

In order to more easily form a limit of relativistic particles in
Section \ref{sec:Relativistic-particles}, we would like to form a
kind of nested worldline gauge theory, with the Galilean spacetime
manifold $\mathcal{G}$ as both the collapsed bundle of a new bundle
and the entire space of the inertial Galilean bundle $(\mathcal{G},T,\pi_{S},S)$.
We accomplish this by defining the base space to again be time $T$,
with the fiber $\mathcal{G}_{t_{0}}$ over time $t_{0}$ a copy of
$\mathcal{G}$, obtaining the \textbf{Galilean nested bundle} $(E_{T}^{\mathcal{G}},T,\pi_{\mathcal{G}},\mathcal{G})$.
At the time $t_{0}$, the value of a matter section $\sigma_{1}$
is defined to be equal to that in $\mathcal{G}$, but in the fiber
$\mathcal{G}_{t_{0}}$; we call this fiber a \textbf{spacetime fiber}
to distinguish it from the space fibers of the Galilean bundle, and
note that this ensures the matter section has one value for each time
coordinate value in $\mathcal{G}$, allowing it to be written 
\begin{equation}
\begin{aligned}x_{1}^{\mu}\left(t\right) & =\left(t,x_{1}^{\mathsf{i}}\left(t\right)\right)\\
\Rightarrow\partial_{t}x_{1}^{\mu} & =\left(1,v_{1}^{\mathsf{i}}\right),
\end{aligned}
\end{equation}
which generalizes to the relativistic setting. Parallel transport
between fibers is defined by an Ehresmann connection which identifies
the points in each copy, and gauge transformations are elements of
the Galilean group applied to all $\mathcal{G}_{t}$ identically,
maintaining a parallel gauge. 

Now, as in Section \ref{subsec:The-Galilean-bundle}, the flat parallel
transport on the fibers $\mathcal{G}_{t}$ means that there is no
uniquely defined metric, and hence we must specify one. We therefore
make this metric part of the state, determined by the choice of gauge
on the inertial Galilean bundle $\mathcal{G}$. A general change of
gauge on $E_{T}^{\mathcal{G}}$ is therefore a combination of two
transformations: a spatial rotation or global displacement applied
to both bundles, and a Galilean boost on $\mathcal{G}$ with a modified
metric to match. As promised in Section \ref{subsec:Geometry}, we
will also see that the choice of metric on $\mathcal{G}$ is indeed
determined by it being a limit of the relativistic metric. 
\begin{figure}[H]
\noindent \begin{centering}
\includegraphics{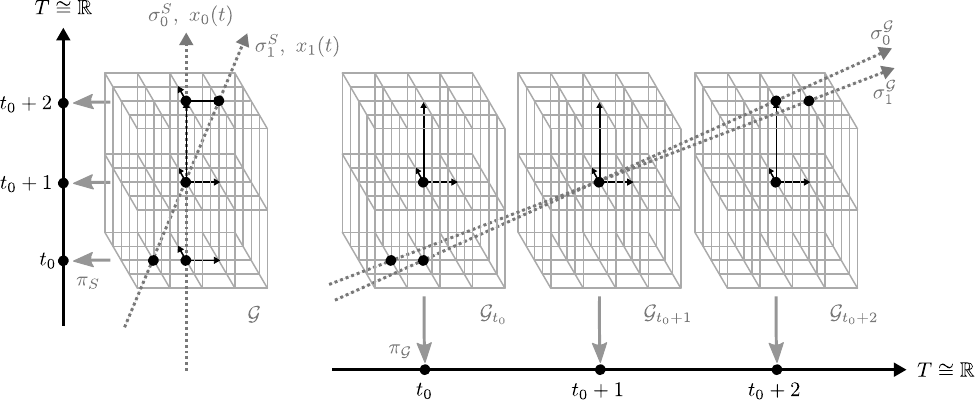}
\par\end{centering}
\caption{\label{fig:The-Galilean-nested-bundle}The Galilean nested bundle
$E_{T}^{\mathcal{G}}$ has a collapsed bundle $\mathcal{G}$ which
is also the entire inertial Galilean bundle, both being called the
Galilean spacetime manifold. At a point $t_{0}$ in time, a matter
section defines a point in the spacetime fiber $\mathcal{G}_{t_{0}}$
which collapses to the matter section over $t_{0}$ in $\mathcal{G}$,
ensuring that the matter section has only one value per time in $\mathcal{G}$.
The metric on the fibers of $E_{T}^{\mathcal{G}}$ are part of the
state, which is determined by Galilean boosts in the underlying bundle
$\mathcal{G}$.}
\end{figure}

\subsection{\label{subsec:PM-action-and-variational-results}The action and equations
of motion}

In the inertial Galilean bundle, the base space is time $T\cong\mathbb{R}$
and a matter section $\sigma_{\alpha}$ is defined for each point
particle indexed by $\alpha$, which in a given gauge (inertial reference
frame) has components $x_{\alpha}^{\mathsf{i}}(t)$ in inertial Galilean
space. The gauge covariant derivative of a matter section in any inertial
frame is $\mathrm{D}_{t}x_{\alpha}^{\mathsf{i}}=\partial_{t}x_{\alpha}^{\mathsf{i}}-v_{\parallel}^{\mathsf{i}}$,
where $v_{\parallel}^{\mathsf{i}}$ is gauge dependent but constant
in time. The metric on $T$ is $h_{tt}=1$, and the spatial metric
is $g_{\mathsf{i}\mathsf{j}}=\delta_{\mathsf{i}\mathsf{j}}$. Despite
these simplifying equalities, we will keep the above general terms
to maintain clarity in object types. 

We now define a force as a spatial vector $F_{\alpha}^{\mathsf{i}}$
associated with each particle at each point in time, defined in each
$S_{t}$ and thus in $S$. We are primarily interested in fundamental
classical forces, which comprise the gravitational and electrostatic
forces. These forces are postulated to be due to two real scalar attributes
associated with each particle: mass $m_{\alpha}$ and charge $q_{\alpha}$.
They are also integrable central forces, i.e. there exists a potential
energy function $U_{12}\left(x_{1}^{\mathsf{i}}-x_{2}^{\mathsf{i}}\right)$
on each $S_{t}$ such that the force between particles is
\begin{equation}
\begin{aligned}F_{12}^{\mathsf{i}} & =-\left(\mathrm{d}U_{12}\right)^{\mathsf{i}}.\end{aligned}
\end{equation}
Specifically, for gravitational and electrostatic forces the potential
energy is 
\begin{equation}
U_{12}=-\frac{k\Upsilon_{1}\Upsilon_{2}}{\left\Vert x_{1}-x_{2}\right\Vert },
\end{equation}
where for gravitation $\Upsilon=m$ and $k=k_{G}$, the gravitational
constant (usually denoted $G$, which we avoid to prevent confusion
with the Einstein tensor), and for electrostatics $\Upsilon=q$ and
$k=k_{e}$, the Coulomb constant. 

For two particles interacting via such a force, the action is defined
to be the time integral of the Lagrangian
\begin{align}
\begin{aligned}L_{\mathrm{PM}U} & =\sum_{\alpha=1,2}\frac{1}{2}m\left\Vert \mathrm{D}_{t}\sigma_{\alpha}\right\Vert ^{2}-U_{12}\\
 & =\sum_{\alpha=1,2}\frac{1}{2}m_{\alpha}h^{tt}g_{\mathsf{ij}}\mathrm{D}_{t}x_{\alpha}^{\mathsf{i}}\mathrm{D}_{t}x_{\alpha}^{\mathsf{j}}-U_{12}(x_{1}-x_{2}),
\end{aligned}
\end{align}
which can be generalized to an arbitrary number of particles. This
Lagrangian results in a generalized momentum canonically conjugate
to $x_{\alpha}$ of
\begin{align}
\begin{aligned}p_{\alpha\mathsf{i}}^{t} & \equiv\frac{\partial L}{\partial\left(\mathrm{D}_{t}x_{\alpha}^{\mathsf{i}}\right)}\\
 & =m_{\alpha}h^{tt}g_{\mathsf{ij}}\mathrm{D}_{t}x_{\alpha}^{\mathsf{j}}\\
 & \equiv g_{\mathsf{ij}}\left(p_{\alpha}^{\mathsf{j}}-m_{\alpha}v_{\parallel}^{\mathsf{j}}\right),
\end{aligned}
\end{align}
where in the last line the (linear) classical momentum 
\begin{equation}
p_{\alpha}^{\mathsf{i}}\equiv m_{\alpha}v_{\alpha}^{\mathsf{i}}
\end{equation}
is distinguished from the canonical momentum $p_{\alpha\mathsf{i}}^{t}$
by the $t$ superscript. Note that unlike $p_{\alpha\mathsf{i}}^{t}$,
$p_{\alpha}^{\mathsf{i}}$ depends upon the inertial reference frame.

The Euler-Lagrange equation for $x_{1}$ is 
\begin{equation}
\begin{aligned}-\frac{\partial U_{12}}{\partial x_{1}^{\mathsf{i}}} & =m_{1}g_{\mathsf{i}\mathsf{j}}\mathrm{D}_{t}\left(v_{1}^{\mathsf{j}}\right)\\
\Rightarrow F_{12}^{\mathsf{i}} & =m_{1}a_{1}^{\mathsf{i}},
\end{aligned}
\end{equation}
which is Newton's second law. The Euler-Lagrange equation for $x_{2}$
will acquire a negative sign, yielding Newton's third law. 

If we vary the Ehresmann connection, which in this case consists of
varying $v_{\parallel}^{\mathsf{i}}$, we get
\begin{align}
\begin{aligned}\sum m_{\alpha}v_{\parallel\mathsf{\mathsf{i}}}-\sum m_{\alpha}v_{\alpha\mathsf{i}} & =0\\
\Rightarrow v_{\parallel}^{\mathsf{i}} & =\frac{1}{\sum m_{\alpha}}\sum m_{\alpha}v_{\alpha}^{\mathsf{i}}\\
 & \equiv v_{\mathrm{CM}}^{\mathsf{i}},
\end{aligned}
\end{align}
so that parallel transport is defined by the center of mass, which
therefore follows an inertial worldline. 

\subsection{\label{subsec:PM-Units}Units}

For particle mechanics including the fundamental gravitational and
electrostatic forces, the base physical dimensions are time $\mathsf{T}$,
length $\mathsf{L}$, mass $\mathsf{M}$, and charge $\mathsf{Q}$.
The units of time are defined by the metric of the base space, while
the units of space are defined by the metric of the fiber in the Galilean
bundle, or by the metric in the space directions in the Galilean spacetime
manifold. In all our geometric gauge theories of particle mechanics,
the scalar particle attributes of mass and charge have no geometric
meaning. 

For the Lagrangian to have consistent units, the potential energy
$U$ must have units of energy $\mathsf{M}\mathsf{L}^{2}\mathsf{T}^{-2}$,
which means that the constants in the definitions of $U$ must have
units

\begin{equation}
\left[k_{G}\right]=\mathsf{M}^{-1}\mathsf{L}^{3}\mathsf{T}^{-2},
\end{equation}
while for the electrostatic force we have 

\begin{equation}
\left[k_{e}\right]=\mathsf{Q}^{-2}\mathsf{M}\mathsf{L}^{3}\mathsf{T}^{-2}.\label{eq:coulomb-const-units}
\end{equation}
For either integrable central force, the action has units of energy-time
\begin{align}
\left[\int\left(\sum_{\alpha=1,2}\frac{1}{2}m_{\alpha}\left\Vert \mathrm{D}_{t}x_{\alpha}\right\Vert ^{2}-U_{12}\right)\mathrm{d}t\right] & =\mathsf{M}\mathsf{L}^{2}\mathsf{T}^{-1}.
\end{align}
These physical dimensions form the foundation of standard dimensionful
units; we will determine their changing geometric meanings in subsequent
physical models as we encounter them.

\section{\label{sec:Relativistic-particles}Relativistic particles}

In defining the theory of relativistic particles, the four-momentum
$P^{\mu}$ is usually presented as the mass $m$ multiplied by a unit
time-like vector $U^{\mu}$, and in geometrized units the mass is
given units of length by defining a physical constant with physical
dimension $\mathsf{L}/\mathsf{M}$ to be unity. Maintaining this
value of unity under a change of units leaves $P^{\mu}$ unchanged
in fixed coordinates, i.e. it is dimensionless; viewing this in a
geometric gauge theory context, $P^{\mu}$ would then seem to be a
tangent vector to the spacetime manifold, i.e. a dimensionless metric-independent
quantity. But its square $g_{\mu\nu}P^{\mu}P^{\nu}=-m^{2}$ is also
assumed to be metric-independent, a contradiction; evidently $P^{\mu}$
is altered by metric variations (to keep $m$ constant) but unchanged
by metric scalings. These odd attributes are accommodated by defining
relativistic particles in terms of an embedded worldline gauge theory. 

\subsection{\label{subsec:The-Minkowski-bundle}The Minkowski bundle}

As mentioned previously, the geometry of relativistic particles is
a generalization of the Galilean nested bundle. Our first modification
is to replace the Galilean spacetime fiber with Minkowski spacetime,
a flat Lorentzian manifold diffeomorphic to $\mathbb{R}^{4}$ which
we denote $\mathbb{M}$ to distinguish from curved spacetime $M$.
As with Galilean space and time metrics, the Lorentzian metric on
$\mathbb{M}$ depends upon a choice of units, and as with the Galilean
spacetime manifold, particles are represented by matter sections $\sigma_{1}$
which in the collapsed bundle are worldlines $x_{1}$. The matter
field is then the gauge covariant derivative of the matter section,
which is denoted 
\begin{equation}
P_{1}\equiv\mathrm{D}x_{1}
\end{equation}
 and which we would like to identify with relativistic four-momentum. 

The relativistic version of an inertial reference frame on $\mathcal{G}$
is a choice of inertial coordinates on $\mathbb{M}$, which are coordinates
whose associated coordinate frame is orthonormal with regard to the
Lorentzian metric, and which is also called an inertial reference
frame. Again treating the manifold point values of $\alpha\colon\mathbb{M}\to\mathbb{R}^{3,1}$
as vectors in the associated vector space, a change of inertial reference
frame, omitting reflections, is a transformation on these inertial
coordinates in the proper orthochronous Poincaré group 
\begin{equation}
ISO(3,1)^{e}\equiv\mathbb{R}^{4}\rtimes SO(3,1)^{e},
\end{equation}
where the superscript denotes the identity component of the Lie group.

With regard to the base space, it cannot be time as in our particle
mechanics geometric gauge theories, since there is no universal time
available. We therefore define the base space to be an arbitrary Riemannian
manifold $\Lambda\cong\mathbb{R}$, with the fiber at each point $\lambda\in\Lambda$
a copy of Minkowski spacetime $\mathbb{M}_{\lambda}$. As with the
Galilean nested bundle, parallel transport is defined by an Ehresmann
connection that identifies the points in each fiber copy, and gauge
transformations are again applied to all fibers at once. In a chosen
gauge, a particle matter section $\sigma_{1}$ projects down to a
worldline $x_{1}^{\mu}\left(\lambda\right)$ in the collapsed bundle
parametrized by $\lambda$ whose gauge covariant derivative is
\begin{equation}
\mathrm{D}_{\lambda}\sigma_{1}^{\mu}=\partial_{\lambda}x_{1}^{\mu}=P_{1}^{\mu}.
\end{equation}

Now, our model thus far includes no relationship between the metric
on the base space and the fibers; but this means that the parametrized
tangent vector of a time-like worldline may vary in magnitude, both
along the worldline and when the metric is varied. This conflicts
with our desire to identify this tangent vector with the four-momentum,
whose magnitude should be constant in both cases. We therefore further
specify our gauge theory as an embedded worldline gauge theory. 

For time-like worldlines, we define the mass to be the coordinate
multiplier 
\begin{equation}
\tau_{1}\equiv\lambda m_{1}
\end{equation}
to the proper time parameter due to the embedding from the parameter
which yields the four-momentum tangent, i.e.

\begin{align}
\begin{aligned}\partial_{\lambda}x_{1}^{\mu} & =m_{1}\partial_{\tau}x_{1}^{\mu}\\
 & \equiv m_{1}U_{1}^{\mu}=P_{1}^{\mu},
\end{aligned}
\end{align}
where $U_{1}$ is defined to be the unit length vector tangent to
the worldline. Thus a time-like particle matter section can be viewed
as a curve $x_{1}^{\mu}\left(\tau\right)$ on the collapsed  bundle
parametrized by proper time and associated with the constant $m_{1}$.
Recall from Section \ref{subsec:Embedded-worldline-gauge-theories}
that $P$ is presumed to be constant under a uniform scaling of the
metric (choice of length unit); thus since the proper time is dependent
upon the choice of units, the mass is as well. Note, however, that
this approach does not work under a change of units which is different
for time and space; this is handled in Section \ref{subsec:Units-RP}. 

We will call this embedded worldline gauge theory $(\mathcal{M},\Lambda,\pi,\mathbb{M})$
the \textbf{Minkowski bundle}, with each fiber $\mathbb{M}_{\lambda}$
called a \textbf{Minkowski fiber}, the base space $\Lambda$ called
the \textbf{parameter space}, and the collapsed  bundle $\mathbb{M}$
referred to as \textbf{Minkowski spacetime}. 
\begin{figure}[H]
\noindent \begin{centering}
\includegraphics{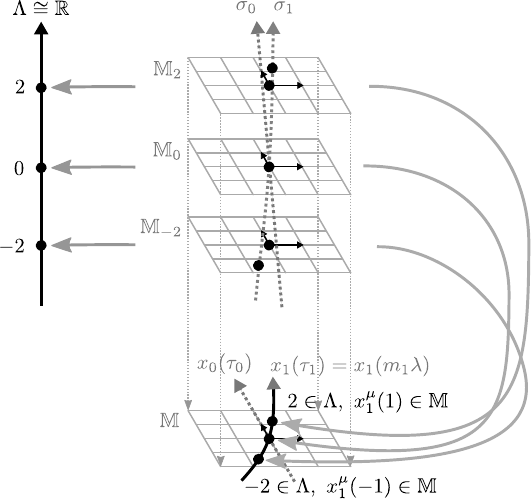}
\par\end{centering}
\caption{\label{fig:The-Minkowski-bundle}The Minkowski bundle is an embedded
worldline gauge theory whose embedded base space for a time-like particle
has an induced proper time metric which is proportionally related
to the non-embedded parameter space $\Lambda$ by a fixed scalar mass
$m_{1}$; in the figure we have $m_{1}=\nicefrac{1}{2}$. This ensures
that parametrized tangents $\partial_{\lambda}x_{1}^{\mu}=m_{1}\partial_{\tau}x_{1}^{\mu}$
have constant magnitude in the Lorentzian metric, thereby manifesting
time dilation. }
\end{figure}

\subsection{\label{subsec:Coordinate-and-metric-dependencies}Coordinate and
metric dependencies}

A gauge on the Minkowski bundle corresponds to a choice of inertial
coordinates, coordinates whose associated coordinate frame is orthonormal
with regard to the fixed Lorentzian metric on $\mathbb{M}$. To position
ourselves for general relativity in Section \ref{sec:General-relativity},
we would like to consider the dependencies of various quantities on
arbitrary coordinate transformations and variations of the metric. 

Since the Minkowski bundle is an embedded worldline gauge theory,
the mass, which is the magnitude of matter section tangents, is not
altered by variations in the metric, but is altered by a uniform scaling
of the metric (choice of length unit). This means that the coordinate-independent
tangent four-momentum $P$ is in fact altered under variation of the
metric, so that $\left\langle P,P\right\rangle =-m^{2}$ remains true,
but is unchanged by a uniform scaling of the metric. In the expression
$P=mU$, $U$ is similarly altered under variation of the metric,
since $\left\langle U,U\right\rangle =-1$ is metric-independent by
definition, and is also changed by a scaling of the metric (choice
of units). We also consider here the particle energy $E$ at a point
$p$, which may geometrically be defined as the time component of
the four-momentum in an orthonormal frame whose time-like basis vector
is parallel to the time coordinate at $p$. Thus $E$ is in general
altered by all of the transformations above, but is not altered by
a change of coordinates which leaves the time-like coordinate parallel
at $p$. Below we list dependencies for all of the basic quantities
associated with the Minkowski bundle.{\footnotesize{}}
\begin{table}[H]
{\footnotesize{}}%
\begin{tabular*}{1\columnwidth}{@{\extracolsep{\fill}}l>{\raggedright}p{0.27\columnwidth}>{\raggedright}p{0.27\columnwidth}>{\raggedright}p{0.27\columnwidth}}
\toprule 
{\footnotesize{}Altered under:} & {\footnotesize{}Metric scalings (length unit)} & {\footnotesize{}Metric variations} & {\footnotesize{}Arbitrary coordinate transformations}\tabularnewline
\midrule 
{\footnotesize{}$m$} & {\footnotesize{}Yes} & {\footnotesize{}No} & {\footnotesize{}No}\tabularnewline
{\footnotesize{}$U$} & {\footnotesize{}Yes} & {\footnotesize{}Yes} & {\footnotesize{}No}\tabularnewline
{\footnotesize{}$U^{\mu}$} & {\footnotesize{}Yes} & {\footnotesize{}Yes} & {\footnotesize{}Yes}\tabularnewline
{\footnotesize{}$P$} & {\footnotesize{}No} & {\footnotesize{}Yes} & {\footnotesize{}No}\tabularnewline
{\footnotesize{}$P^{\mu}$} & {\footnotesize{}No} & {\footnotesize{}Yes} & {\footnotesize{}Yes}\tabularnewline
{\footnotesize{}$E$} & {\footnotesize{}Yes} & {\footnotesize{}Yes} & {\footnotesize{}Yes}\tabularnewline
\bottomrule
\end{tabular*}{\footnotesize\par}

{\footnotesize{}\caption{Alteration of basic quantities associated with the Minkowski bundle
under a change of length unit, variations of the metric, and arbitrary
changes of curvilinear coordinates. A uniform scaling of the metric
changes $m$ and $U$, but not their product $P=mU$ and therefore
not the components $P^{\mu}$ in fixed coordinates. The component
$E$ however is defined to be the time component of $P$ in an orthonormal
frame, which does change with units, since the four-vector $P$ does
not. Like $P$, however, $E$ is altered by non units related metric
variations, since the length of $P$ can change due to metric changes
which are orthogonal to the time coordinate axis, thus changing its
time component. $E$ can also be altered by coordinate transformations
which change the direction of the time coordinate.}
}{\footnotesize\par}
\end{table}
The invariance properties of the basic quantities above are identical
in the usual treatments of relativistic physics in geometrized units,
but are usually not explicitly mentioned nor associated with any geometric
model. 

\subsection{\label{subsec:RP-action}The action and equations of motion}

Four-momentum must be conserved in order to obtain a relativistic
theory which yields particle mechanics in the non-relativistic limit.
This requirement however, ends up making particle interactions impossible.
One can see this by noting that in Minkowski spacetime there is no
reference frame independent concept of simultaneity, and so no way
to define a reference frame independent total four-momentum which
can be constant in time. This inability to consistently define interactions
between relativistic particles is sometimes called the no-interaction
theorem, and can be shown to be true under a number of different basic
assumptions (see \cite{Ekstein} for a general approach to the topic
which includes both quantum and classical particles).

We therefore address a single free time-like particle, and drop identifying
subscripts for all variables except $x_{1}$ to avoid confusion with
coordinates. The action is defined to be

\begin{equation}
\begin{aligned}S_{\mathrm{TP}} & \equiv-m\int\left\Vert \mathrm{D}_{\lambda}\sigma\right\Vert \mathrm{d}\lambda\\
 & =-m\int\sqrt{-\left\langle \mathrm{D}_{\lambda}x_{1}^{\mu},\mathrm{D}_{\lambda}x_{1}^{\mu}\right\rangle }\mathrm{d}\lambda,
\end{aligned}
\end{equation}
which is proportional to the arc length of the worldline in the Minkowski
metric, and is hence independent of the parametrization, so that we
may choose coordinates on Minkowski spacetime and then change variables
to set the parameter equal to the time coordinate. In dimensionful
units, we find in Section \ref{subsec:Units-RP} that

\begin{equation}
\begin{aligned}S_{\mathrm{TP}} & =-mc\int\sqrt{-\left\langle \partial_{t}x_{1}^{\mu},\partial_{t}x_{1}^{\mu}\right\rangle }\mathrm{d}t\\
 & \sim-mc\int\left(c-\frac{1}{2}\frac{v^{2}}{c}\right)\mathrm{d}t\\
 & =\int\left(\frac{1}{2}mv^{2}-mc^{2}\right)\mathrm{d}t
\end{aligned}
\end{equation}
where the second line we use a Taylor expansion to order $v^{2}$.
This confirms that our action matches that of particle mechanics in
the limit $v<<c$, since the constant ``rest energy'' term $-mc^{2}$
does not affect the EOM; in this same limit, the transformation group
$ISO(3,1)^{e}$ becomes $\mathrm{SGal}(3)$.

Note that we may define a constant Lagrangian which is numerically
identical to the above, and by using $t=\gamma\lambda m$ we can see
also has the same non-relativistic limit in dimensionful units:

\begin{equation}
\begin{aligned}S_{\mathrm{\emptyset P}} & \equiv-\int m^{2}c^{2}\mathrm{d}\lambda.\end{aligned}
\end{equation}
Reverting to geometric units, this enables us to construct an action
with the same non-relativistic limit, and which positions us for the
transition to relativistic continua when reparametrized by proper
time:

\begin{equation}
\begin{aligned}S_{\mathrm{TP\emptyset}} & \equiv\frac{1}{2}\int\left(-m\left\Vert \mathrm{D}_{\lambda}\sigma\right\Vert -m^{2}\right)\mathrm{d}\lambda\\
 & =\frac{1}{2}\int\left(-m\left\Vert \mathrm{D}_{\tau}\sigma\right\Vert -m\right)\mathrm{d}\tau.
\end{aligned}
\end{equation}
For any of the above actions, the Euler-Lagrange equations result
in the EOM 
\begin{align}
\begin{aligned}0 & =\mathrm{D}_{\lambda}\left(g_{\mu\nu}P^{\nu}\right)\\
\Rightarrow\partial_{\lambda}P^{\mu} & =0,
\end{aligned}
\end{align}
i.e. the tangent to the worldline is its own parallel transport, and
hence the worldline is straight. The form of the Lagrangian also
requires the worldline to be time-like, and the embedding does the
same as well as requiring $P$ to be future-directed.

\subsection{\label{subsec:Units-RP}Units}

Unlike the Galilean nested bundle, the Minkowski bundle features a
fiber with a fixed metric, which we can use to define all physical
dimensions in geometric units. The four-momentum $P$ is geometrically
the tangent vector to the embedded base space $\Lambda$ parametrized
by a coordinate $\lambda$ on $\Lambda$ whose coordinate frame has
unit length. Recall from Section \ref{subsec:The-Minkowski-bundle}
that $P$ is dimensionless with regard to these units, so that
\begin{equation}
\begin{aligned}\left[P^{\mu}\partial_{\mu}\right]_{\mathrm{G}} & =\mathsf{1},\\
\Rightarrow\left[P^{\mu}\right]_{\mathrm{G}} & =\mathsf{L},
\end{aligned}
\end{equation}
where we recall that $\left[\partial_{\mu}\right]_{\mathrm{G}}=\mathsf{L}^{-1}$.
Since $U^{\mu}$ and the metric components $g_{\mu\nu}$ are dimensionless,
we then have
\[
\begin{aligned}\left[m\right]_{\mathrm{G}} & =\mathsf{L}.\end{aligned}
\]
For a time-like particle, at any point the proper time $\tau$ is
a unit of length on $\mathbb{M}$ parallel to $P$; thus $\begin{aligned}\left[\tau\right]_{\mathrm{G}} & =\mathsf{L}\end{aligned}
$, and since $\tau=\lambda m$, we have 
\begin{equation}
\begin{aligned}\left[\lambda\right]_{\mathrm{G}} & =\mathsf{1}.\end{aligned}
\end{equation}
We can view $m$ in geometric units as ``the length in $\mathbb{M}$
of a unit distance on $\Lambda$'' due to its embedding in $\mathbb{M}$,
and the dimensionless $\lambda$ can be viewed as ``the ratio of
the length of unit time to the length of unit mass.'' We will arrive
at a different geometric meaning of mass in Section \ref{sec:General-relativity}. 

We can also see that in geometric units our actions all have the same
physical dimension
\begin{equation}
\left[\int m\sqrt{-\left\langle \mathrm{D}_{\lambda}x_{1}^{\mu},\mathrm{D}_{\lambda}x_{1}^{\mu}\right\rangle }\mathrm{d}\lambda\right]_{\mathrm{G}}=\left[\int m^{2}\mathrm{d}\lambda\right]_{\mathrm{G}}=\mathsf{L}^{2}.
\end{equation}
Reparametrizing these actions does not change the physical dimension,
e.g. since $\partial_{\tau}x_{1}^{\mu}=U_{1}^{\mu}$ is dimensionless
we have 

\begin{equation}
\left[\frac{1}{2}\int\left(-m\left\Vert \mathrm{D}_{\tau}x_{1}\right\Vert -m\right)\mathrm{d}\tau\right]=\mathsf{L}^{2}.
\end{equation}

We would now like to relate the above physical dimensions to those
in dimensionful units, specifically SI units. Since geometric units
are unique to geometric gauge theory, we treat this topic in some
detail. We first address the speed of light $c$, which is the ratio
of a length in a time-like direction to a length in a space-like direction,
and is therefore dimensionless in geometric units, with a value of
unity. In SI units, we choose to view length in time-like directions
as a different base physical dimension $\mathsf{T}$, i.e. 
\begin{equation}
\left[c\right]_{\mathrm{G}}=\mathsf{1},\qquad\left[c\right]_{\mathrm{SI}}=\mathsf{L}\mathsf{T}^{-1}.
\end{equation}
Geometrically, we can view $c$ as ``the length of a unit of time.''
We can convert expressions in geometric units to the equivalent expression
in SI units by inserting a $c$ for all time-like lengths, e.g.
\begin{equation}
\tau_{\mathrm{G}}\rightarrow c\tau_{\mathrm{SI}},
\end{equation}
where $\left[\tau_{\mathrm{G}}\right]=\mathsf{L}$ and $\left[\tau_{\mathrm{SI}}\right]=\mathsf{T}$. 

Moving on to the four-velocity $\partial_{\tau}x_{1}^{\mu}=U^{\mu}$,
we have 
\begin{equation}
\begin{aligned}U_{\mathrm{G}}^{\mu} & \rightarrow c^{-1}U_{\mathrm{SI}}^{\mu}\\
\Rightarrow\left\langle U,U\right\rangle _{\mathrm{SI}} & =c^{2}.
\end{aligned}
\end{equation}
Thus in geometric units $U_{\mathrm{G}}^{\mu}$ are the dimensionless
components of a unit time-like vector parallel to the worldline, but
in dimensionful units $U_{\mathrm{SI}}^{\mu}$ are components of a
time-like vector of length $c$ parallel to the worldline with physical
dimension of velocity.

We next address mass, which in SI units we choose to treat as a different
base physical dimension $\mathsf{M}$. Including $c$, our proportionality
relation is $c\tau=\lambda m$, so that 
\begin{equation}
\left[\lambda\right]_{\mathrm{G}}=\mathsf{1},\qquad\left[\lambda\right]_{\mathrm{SI}}=\mathsf{M}^{-1}\mathsf{L},
\end{equation}
and geometrically, we can view $\lambda$ as ``the geometric length
of a unit of mass.'' In SI units this is written as 
\begin{align}
\begin{aligned}\lambda & \equiv16\pi k_{G}c^{-2}\\
\Rightarrow\left[k_{G}\right]_{\mathrm{SI}} & =\mathsf{M}^{-1}\mathsf{L}^{3}\mathsf{T}^{-2},
\end{aligned}
\end{align}
which is defined in order to obtain $k_{G}$ as the gravitation constant
in the Newtonian limit as we will see in Section \ref{subsec:Units-GR}.
We can convert expressions in geometric units to the equivalent expression
in SI units by replacing all masses, i.e.
\begin{equation}
\begin{aligned}m_{\mathrm{G}} & \rightarrow16\pi k_{G}c^{-2}m_{\mathrm{SI}},\end{aligned}
\label{eq:mass-conversion}
\end{equation}
where $\left[m_{\mathrm{G}}\right]=\mathsf{L}$ and $\left[m_{\mathrm{SI}}\right]=\mathsf{M}$.

Applying our results to e.g. our action in terms of proper time, we
have 

\begin{equation}
\begin{aligned}\frac{1}{2}\int\left(-m\left\Vert \mathrm{D}_{\tau}x_{1}\right\Vert -m\right)\mathrm{d}\tau & \rightarrow8\pi k_{G}c^{-2}\int\left(-mc^{-1}\left\Vert U\right\Vert -m\right)c\mathrm{d}\tau,\\
\left[8\pi k_{G}c^{-1}\int\left(-mc^{-1}\left\Vert U\right\Vert -m\right)\mathrm{d}\tau\right]_{\mathsf{SI}} & =\mathsf{L}^{2}.
\end{aligned}
\end{equation}
The convention, however, is to present the action in SI units of energy-time,
to match that of particle mechanics. To accomplish this, we introduce
another conversion factor
\begin{equation}
\begin{aligned}S_{\mathrm{G}} & \rightarrow\frac{c^{3}S_{\mathsf{SI}}}{16\pi k_{G}}\\
\Rightarrow\left[S_{\mathsf{SI}}\right] & =\mathsf{M}\mathsf{L}^{2}\mathsf{T}^{-1},
\end{aligned}
\end{equation}
which for action terms with a single mass factor can be combined to
yield
\begin{equation}
\begin{aligned}m_{\mathrm{G}} & \overset{S\propto m}{\rightarrow}cm_{\mathrm{SI}}.\end{aligned}
\end{equation}

These conversions then result in

\begin{equation}
\begin{aligned}\frac{1}{2}\int\left(-m\left\Vert \mathrm{D}_{\tau}x_{1}\right\Vert -m\right)\mathrm{d}\tau & \rightarrow\frac{1}{2}\int\left(-mc\left\Vert U\right\Vert -mc^{2}\right)\mathrm{d}\tau,\end{aligned}
\end{equation}
which was used in Section \ref{subsec:RP-action}. {\footnotesize{}}{\footnotesize\par}

\section{\label{sec:Relativistic-continua}Relativistic continua}

The theory of relativistic continua is expressed in terms of the four-current,
yet it is the underlying worldlines which are varied to obtain EOM.
Here we present a modern geometric take on Dirac's approach to this
topic in \cite{Dirac}. A reference to a complementary derivation
which is less geometric and more rigorous would be welcome, but seems
to not be readily found. As in Section \ref{subsec:Coordinate-and-metric-dependencies},
to prepare for curved spacetime we will track whether the quantities
we consider are invariant under (non units related) variations of
the metric and (arbitrary curvilinear) coordinate transformations;
we will also consider (inertial coordinate) Lorentz transformations
and include the factor $\sqrt{g}\equiv\sqrt{\left|\det\left(g_{\mu\nu}\right)\right|}$
despite its value of unity in flat spacetime. 

\subsection{\label{subsec:The-Minkowski-bundle-congruence}The Minkowski bundle
congruence}

Starting from the worldlines of relativistic particles, we would like
to construct a quantity which has a continuous value at every point
in spacetime. To do so, we first consider a congruence of curves on
$\mathbb{M}$, the flow of a non-vanishing smooth vector field. In
our context, this can be viewed as an infinite family of non-intersecting
matter sections $\sigma_{\alpha}$ of the Minkowski bundle $\mathcal{M}$
which fill Minkowski spacetime $\mathbb{M}$ and vary continuously,
each of which corresponds to an embedding of $\Lambda$ in $\mathbb{M}$.
We furthermore view the congruence as a limit where, as the number
of worldlines approaches infinity, an infinitesimal spacetime hyperplane
contains an infinitesimal amount of a non-negative real quantity called
the particle number $N$, which is proportional to the volume of the
infinitesimal hyperplane. We will call this structure a \textbf{worldline
congruence}. Note that the limit taken to arrive at a worldline congruence
preserves the idea of a density of worldlines, which is absent in
the usual congruence of curves concept used in general relativity.

Under this limit, quantities associated with each worldline become
quantities per particle number. In particular, we maintain the association
of a rest mass $m$ with each time-like worldline, giving the length
of a unit distance on the associated embedded $\Lambda$; we also
assume that this rest mass is identical for all worldlines, since
any variation can instead be captured by the number of the worldlines
themselves. In the limit of infinite worldlines, the mass per worldline
then becomes a constant rest mass $m$ per particle number.  

We will call this entire structure, which is a geometrized definition
of relativistic dust, a \textbf{Minkowski bundle congruence}; each
fiber is again called a Minkowski fiber, the base space is called
parameter space, and the collapsed bundle is called Minkowski spacetime.
In this structure, the intrinsic objects may be viewed as the worldline
curves $x_{\alpha}$, the infinitesimal particle number per unit space-like
hyperplane at each point, and a single mass parameter $m$ per particle
number. 
\begin{figure}[H]
\noindent \begin{centering}
\includegraphics{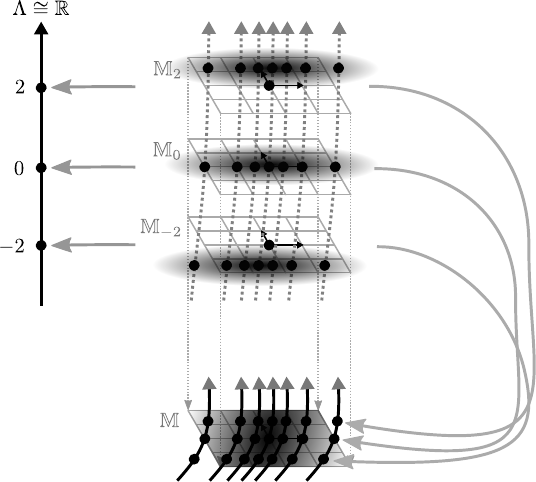}
\par\end{centering}
\caption{The Minkowski bundle congruence is a congruence of matter field curves,
each of which is associated with a single embedded worldline gauge
theory; the density of section values on each $M_{\lambda}$ is depicted
by a gradient, as is the density of worldlines on $M$. For a time-like
congruence, the induced proper time metric is proportionally related
to the embedded parameter space $\Lambda$ by a fixed scalar mass
$m$ for all worldlines. In the limit of infinite worldlines, any
infinitesimal space-like hyperplane in $\mathbb{M}$ is associated
with an infinitesimal particle number $N$ which is proportional to
the volume of the hyperplane.}
\end{figure}

\subsection{\label{subsec:Currents-and-densities}Currents, densities, and the
SEM tensor}

We now may define the (particle number) current pseudo 3-form
\begin{equation}
\xi\left(u,v,w\right)\equiv\frac{\mathrm{d}N}{\mathrm{d}V\left(u,v,w\right)},
\end{equation}
whose value at each point $p$ is the infinitesimal particle number
per unit hypersurface defined by its argument vectors. Introducing
metric dependence, we use the hodge star and index raising to define
the four-current 
\begin{align}
\begin{aligned}J & \equiv\left(*\xi\right)^{\sharp}\\
 & \equiv\rho_{0}U,
\end{aligned}
\end{align}
a four-vector which may be expressed at each point in terms of the
unit vector $U$ parallel to the worldline and the rest density $\rho_{0}$,
which is the worldline rest frame particle number per unit volume
(or the infinitesimal particle number per unit space-like hypersurface
orthogonal to $U$).

We can exchange metric dependence for coordinate dependence by defining
the \textbf{four-current density} as
\begin{align}
\begin{aligned}\mathfrak{J} & \equiv J\sqrt{g}\\
 & \equiv\breve{\rho}_{0}\breve{U},
\end{aligned}
\end{align}
where the \textbf{coordinate density} $\breve{\rho}_{0}$ can be seen
to be the infinitesimal particle number per unit coordinate space-like
hypersurface, and the \textbf{coordinate unit tangent}
\begin{equation}
\breve{U}\equiv\frac{U}{U^{0}}
\end{equation}
is the worldline tangent vector with unit time component in the chosen
arbitrary coordinates. Note that in inertial coordinates we have $\mathfrak{J}=J$,
which remains invariant under Lorentz transformations. Since our worldline
congruence is constructed from the limit of unbroken worldlines, an
equal particle number enters and exits any arbitrary volume of spacetime,
so that
\begin{equation}
\begin{aligned}\mathrm{div}\left(J\right) & =-*\mathrm{d}\xi=\partial_{\mu}\mathfrak{J}^{\mu}=0.\end{aligned}
\end{equation}

We construct a general stress energy momentum tensor (SEM tensor)
from multiple time-like dust four-currents
\begin{equation}
\begin{aligned}T^{\mu\nu} & \equiv\sum_{\alpha}T_{\mathrm{TD}\alpha}^{\mu\nu}\\
 & \equiv\sum_{\alpha}P_{\alpha}^{\mu}\otimes J_{\alpha}^{\nu}\\
 & =m\sum_{\alpha}\rho_{0\alpha}U_{\alpha}^{\mu}U_{\alpha}^{\nu},
\end{aligned}
\end{equation}
with the corresponding SEM tensor density defined as
\begin{align}
\mathfrak{T}^{\mu\nu} & \equiv\sum_{\alpha}P_{\alpha}^{\mu}\otimes\mathfrak{J}_{\alpha}^{\nu}\\
 & =\sqrt{g}T^{\mu\nu},
\end{align}
which unlike $\mathfrak{J}$ remains metric-dependent due to the $P=mU$
component. 

\subsection{\label{subsec:Worldline-variation}Worldline variation}

Our actions will be expressed in terms of four-currents, but we will
be varying the matter sections of the Minkowski bundles which define
the worldline congruence. We must therefore determine how such a variation
affects $J$; in order to maintain validity in general relativity,
we do so on a curved spacetime manifold $M$. What follows is a modern
geometric take on Dirac's approach in \cite{Dirac} pp. 51-52. 

We accomplish a smooth worldline congruence variation by transforming
each worldline $x_{\alpha}$ to $x_{\alpha}^{\prime}=x_{\alpha}+\varepsilon b_{\alpha}$,
where $b$ is a smooth vector field. The tangent vector $J$ at a
point $p$ is then transported by the flow of the vector field $b$.
So if $\Phi_{\xi}$ is the local one-parameter group of diffeomorphisms
associated with $b$, we have
\begin{align}
\begin{aligned}\left.J^{\prime}\right|_{p+\varepsilon b} & \propto\mathrm{d}\Phi_{\varepsilon}\left(\left.J\right|_{p}\right)\\
\Rightarrow J^{\prime} & \propto J-\varepsilon L_{b}J,
\end{aligned}
\end{align}
where we have recalled the definition of the Lie derivative (see e.g.
\cite{Marsh-book} pp. 97-98) 
\begin{align}
L_{b}J & =\underset{\varepsilon\rightarrow0}{\textrm{lim}}\frac{1}{\varepsilon}\left[J\left|_{p+\varepsilon b}\right.-\mathrm{d}\Phi_{\varepsilon}\left(J\left|_{p}\right.\right)\right].
\end{align}

\begin{figure}[H]
\noindent \begin{centering}
\includegraphics{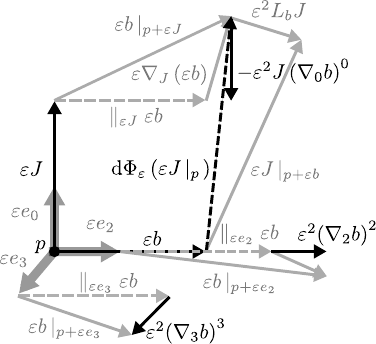}
\par\end{centering}
\caption{\label{fig:field-current-variation}Without loss of generality, we
define the time-like basis vector $e_{0}$ to be parallel to $J$,
and $e_{2}$ to be parallel to the variation $b$ (any component parallel
to $J$ will not affect either the direction or density of worldlines).
The flow of $b$ takes the vector $\varepsilon J$ to $\mathrm{d}\Phi_{\varepsilon}\left(\varepsilon\left.J\right|_{p}\right)$,
with the difference from the value of $\varepsilon J$ at that point
being $-\varepsilon^{2}L_{b}J$. This establishes the direction $U^{\prime}$
of the new four-current, but in order to calculate its length we first
normalize it to the original length $\varepsilon\rho_{0}$ by subtracting
the extra length created by the flow, which in terms of the Levi-Civita
connection is $\varepsilon^{2}J\left(\nabla_{0}b\right)^{0}$. The
change in length is then the change in density; the change in density
due to the hyperplane being orthogonal to a new direction is of higher
order in $\varepsilon$, so our last step is to find the new worldline
density $\varepsilon\rho_{0}^{\prime}$ due to the factor by which
surrounding worldlines are spread out by the flow of $b$. To order
$\varepsilon^{2}$, this is determined by the other components of
the divergence $\varepsilon^{2}J\nabla_{\mu}b^{\mu}$: if e.g. the
length $\varepsilon e_{3}$ originally contained $N$ worldlines,
this number will be reduced by $\varepsilon^{2}N\left(\nabla_{3}b\right)^{3}$;
therefore, the new worldline density is $\varepsilon\rho_{0}^{\prime}=\varepsilon\rho_{0}-\sum\varepsilon^{2}\rho_{0}\left(\nabla_{\mathsf{i}}b\right)^{\mathsf{i}}$,
and the total new four-current is $\varepsilon J^{\prime}=\varepsilon J-\varepsilon^{2}L_{b}J-\varepsilon^{2}J\nabla_{\mu}b^{\mu}$.}
\end{figure}

To determine the length of $J^{\prime}$, we must first subtract the
additional length added by the flow of $b$, since this length change
represents the change in the particle number density $\rho_{0}$,
which must be calculated separately. As depicted in Figure \ref{fig:field-current-variation},
this additional length is the component of $\varepsilon\nabla_{J}b$
parallel to $J$, where the covariant derivative $\nabla$ is that
of the Levi-Civita connection of the metric on $M$.

With the length of $J^{\prime}$ thus matched to the original length,
we must then account for what does result in a change in $\rho_{0}$,
the particle number per orthogonal space-like hypersurface: the expansion
due to the flow of $b$, which changes the distance to the surrounding
worldlines after being transported. Again referring to Figure \ref{fig:field-current-variation},
this factor applied to $J$ is the sum of the components of the negative
divergence of $b$ orthogonal to $J$, which added to the previous
component yields the total divergence. We therefore have
\begin{align}
\begin{aligned}J^{\prime} & =J-\varepsilon L_{b}J-\varepsilon J\mathrm{div}\left(b\right)\\
\Rightarrow\delta J & =-L_{b}J-J\mathrm{div}\left(b\right),
\end{aligned}
\end{align}
which in this form is valid in the presence of both curvature and
torsion.

By writing the Lie derivative in terms of the covariant derivative,
we can express this in arbitrary coordinates as

\begin{align}
\delta J^{\mu} & =\nabla_{\nu}\left(J^{\nu}b^{\mu}-b^{\nu}J^{\mu}\right),
\end{align}
which makes it manifest that the variation vanishes if $b$ is parallel
to $J$ and that $\nabla_{\mu}\left(\delta J^{\mu}\right)=0$. It
is not hard to show that the four-current density has the similar
variation

\begin{align}
\begin{aligned}\delta\mathfrak{J}^{\mu} & =\partial_{\nu}\left(\mathfrak{J}^{\nu}b^{\mu}-b^{\nu}\mathfrak{J}^{\mu}\right).\end{aligned}
\end{align}

\subsection{\label{subsec:RD-action}The action and equations of motion}

To define an action for our worldline congruence which results in
EOM which are consistent with those of the relativistic particle gauge
theory associated with each worldline, we replace the integrand per
unit length $L_{\tau}\mathrm{d}\tau$ along a single worldline by
a 4-form $L_{\mathrm{D}}\mathrm{d}V$, which applied to an orthonormal
basis yields this quantity multiplied by the particle number per unit
space-like hypersurface orthogonal to the worldline, i.e. at each
point we define 

\begin{equation}
\begin{aligned}L_{\mathrm{D}}\mathrm{d}V & \equiv L_{\tau}\rho_{0}\mathrm{d}V.\end{aligned}
\end{equation}
Applying this to our relativistic particle action $S_{\mathrm{TP\emptyset}}$
yields
\begin{equation}
\begin{aligned}S_{\mathrm{TD}} & \equiv\frac{1}{2}\rho_{0}\int\left(-m\left\Vert \mathrm{D}_{\tau}\sigma\right\Vert -m\right)\mathrm{d}V\\
 & =-m\int\rho_{0}\mathrm{d}V\\
 & =-m\int\sqrt{-\left\langle J,J\right\rangle }\mathrm{d}V.
\end{aligned}
\end{equation}

We derive the Euler-Lagrange equation for a variation of the worldlines
$x_{\alpha}$ in some detail, again expanding on Dirac's treatment
in \cite{Dirac}: 
\begin{equation}
\begin{aligned}\delta S_{\mathrm{TD}}\left(x_{\alpha}^{\mu}\right) & =-m\int\frac{-J_{\mu}}{\sqrt{-\left\langle J,J\right\rangle }}\delta J^{\mu}\mathrm{d}V\\
 & =m\int U_{\mu}\nabla_{\nu}\left(J^{\nu}b^{\mu}-b^{\nu}J^{\mu}\right)\mathrm{d}V\\
 & =-m\int\left(J^{\nu}b^{\mu}-b^{\nu}J^{\mu}\right)\nabla_{\nu}U_{\mu}\mathrm{d}V\\
 & =m\int\left(\nabla_{\nu}U_{\mu}-\nabla_{\mu}U_{\nu}\right)b^{\nu}J^{\mu}\mathrm{d}V\\
 & =m\int\left(\left\langle \nabla_{b}U,J\right\rangle -\left\langle \nabla_{J}U,b\right\rangle \right)\mathrm{d}V\\
 & =-m\int\left\langle \rho_{0}\nabla_{U}U,b\right\rangle \mathrm{d}V=0\\
\Rightarrow\nabla_{U}U & =0.
\end{aligned}
\end{equation}
In the second and penultimate lines we use $J=\rho_{0}U$, in the
third line we integrate by parts and use the divergence theorem recalling
that $b$ vanishes on the boundary, in the fourth line we relabel
dummy indices, and in the penultimate line we use the fact that since
$U$ is of constant length its covariant derivative cannot have any
component in the $U$ direction, which is parallel to $J$. We get
the same result if we vary $\mathfrak{J}$.

Thus, as in the case of relativistic particles, the equation of motion
restrict worldlines to ``straight lines,'' in this case the geodesics
of the Levi-Civita connection. Note that since the Levi-Civita connection
is also used in the expression used with the divergence theorem, this
analysis continues to apply in the presence of both curvature and
torsion, i.e. in the presence of torsion the worldlines continue to
follow Levi-Civita (metric) geodesics.

\subsection{\label{subsec:Units-RD}Units}

The number of particles is a dimensionless quantity, and therefore
since the rest density is the number of particles per unit space-like
volume, we have
\begin{equation}
\begin{aligned}\left[\rho_{0}\right]_{\mathrm{G}} & =\left[J^{\mu}\right]_{\mathrm{G}}=\mathsf{L}^{-3}.\end{aligned}
\end{equation}
In dimensionful units the conversion factor $c^{-1}$ in $U^{\mu}$
results in
\begin{equation}
\begin{aligned}J^{\mu} & \rightarrow c^{-1}J^{\mu}\\
\Rightarrow\left[J^{\mu}\right]_{\mathrm{SI}} & =\mathsf{L}^{-2}\mathsf{T}^{-1},
\end{aligned}
\end{equation}
whose components are $\left(c\rho_{0},j^{\mathsf{i}}\right)$, preserving
the definition of $\rho_{0}$ in both units as the number of particles
per volume, while matching the classical definition of the current
$j$ as the particles per second passing through the unit area perpendicular
to their direction.

Recalling the attributes of four-momentum from Section \ref{subsec:Units-RP},
for the SEM tensor we have
\begin{align}
\begin{aligned}\left[T^{\mu\nu}\right]_{\mathrm{G}} & =\mathsf{L}^{-2},\\
T^{\mu\nu} & \rightarrow16\pi k_{G}c^{-4}T^{\mu\nu},\\
\left[T^{\mu\nu}\right]_{\mathrm{SI}} & =\mathsf{M}\mathsf{L}^{-1}\mathsf{T}^{-2},
\end{aligned}
\end{align}
which in SI units is that of energy density or momentum current. 

The volume element now has physical dimension 
\[
\begin{aligned}\left[\mathrm{d}V\right]_{\mathrm{G}} & =\mathsf{L}^{4},\end{aligned}
\]
so that as with relativistic particles, the action in geometric units
is 
\[
\begin{aligned}\left[-m\int\sqrt{-\left\langle J,J\right\rangle }\mathrm{d}V\right]_{\mathrm{G}} & =\mathsf{L}^{2}.\end{aligned}
\]
In SI units we have
\[
\begin{aligned}\left[\mathrm{d}V\right]_{\mathrm{SI}} & =\mathsf{L}^{3}\mathsf{T}\\
\Rightarrow\mathrm{d}V & \rightarrow c\mathrm{d}V,
\end{aligned}
\]
and using our conversion factors here and from Section \ref{subsec:Units-RP},
we obtain actions maintaining units of energy-time:

\[
\begin{aligned}-m\int\sqrt{-\left\langle J,J\right\rangle }\mathrm{d}V & \rightarrow-mc\int\sqrt{-\left\langle J,J\right\rangle }\mathrm{d}V,\\
-m\int\rho_{0}\mathrm{d}V & \rightarrow-mc^{2}\int\rho_{0}\mathrm{d}V,\\
\left[-mc^{2}\int\rho_{0}\mathrm{d}V\right]_{\mathrm{SI}} & =\mathsf{M}\mathsf{L}^{2}\mathsf{T}^{-1}=\left[-mc\int\sqrt{-\left\langle J,J\right\rangle }\mathrm{d}V\right]_{\mathrm{SI}}.
\end{aligned}
\]

\section{\label{sec:General-relativity}General relativity}

With the preparatory work of the last two sections, the most straightforward
geometry for general relativity consists of mainly replacing the flat
parallel transport on Minkowski spacetime with an independent quantity
which is part of the state. However, a common desire is to view the
spacetime connection as that of a sort of gauge theory, usually at
the cost of stretched definitions. Using geometric gauge theory, we
may accommodate this view in an arguably more natural fashion that
includes a matter field; but more importantly, this approach ends
up being similar to that of the geometric gauge theory of electromagnetism,
providing a straightforward transition between the two theories.

\subsection{The Einstein bundle congruence}

Our first geometry is obtained by modifying the Minkowski bundle
such that the fibers, now denoted $M_{\lambda}$, and the collapsed
worldline bundle $M$, are all copies of a curved spacetime Lorentzian
manifold, whose otherwise arbitrary parallel transport is part of
the state. We assume $M$ is orientable and time-orientable. We call
this the \textbf{Einstein bundle}\index{Einstein bundle} $(\mathcal{E},\Lambda,\pi,M)$
(some authors may use this term differently). 

We can now follow the same approach as in the last section, and form
a worldline congruence of Einstein bundle base spaces, with each particle
number associated with a fixed mass, a choice of parallel gauge a
choice of coordinates (now arbitrary and only local), and variations
of the matter sections required to be continuous so that they amount
to a variation of the worldlines to form a new congruence. We will
call this structure an \textbf{Einstein bundle congruence}\index{Einstein bundle congruence},
and also as in the last chapter we consider the fundamental geometric
objects to be the worldline curves $x_{\alpha}$ and the infinitesimal
particle number per unit space-like hyperplane $\rho_{0}$ at each
point.

\subsection{The Einstein nested bundle }

In order to view the spacetime connection as that of a matter field,
we define an embedded geometric gauge theory as described in Section
\ref{subsec:Embedded-spacetime-gauge-theories}, whose base space
is the spacetime manifold $M$, and whose fiber $X_{p}$ over every
point $p\in M$ is isometric to $M$. The matter section over $p$,
defined to be parallel according to a flat Ehresmann connection, is
just the corresponding point in $X_{p}$. A matter field is a time-like
future-directed unit length covector $\hat{\underline{\Phi}}$ on
each fiber, which collapses to a differential 1-form on $M$ also
denoted $\hat{\underline{\Phi}}$, and whose parallel transport is
that of the spacetime connection on $M$. The length 
\begin{equation}
\left\Vert \hat{\underline{\Phi}}\right\Vert =\sqrt{-g^{\mu\nu}\hat{\underline{\Phi}}_{\mu}\hat{\underline{\Phi}_{\nu}}}=1
\end{equation}
is therefore calculated using the metric on $M$, and is not invariant
under metric variations. The matter field is thus a section of the
cotangent bundle $(T^{*}M,M,\pi_{M},\mathbb{R}^{3,1})$, and the parallel
transport of a matter field has a connection with values in $so(3,1)$
in an orthonormal coframe. 

We then define $M$ to also be the collapsed bundle of the Einstein
bundle congruence $(\mathcal{E},\Lambda,\pi_{\Lambda},M)$; we thus
construct a ``nested'' geometric gauge theory $(T^{*}M_{\mathcal{E}},M,\pi_{M},\mathbb{R}^{3,1})$,
which we call the\textbf{ Einstein nested bundle}. 
\begin{figure}[H]
\noindent \begin{centering}
\includegraphics{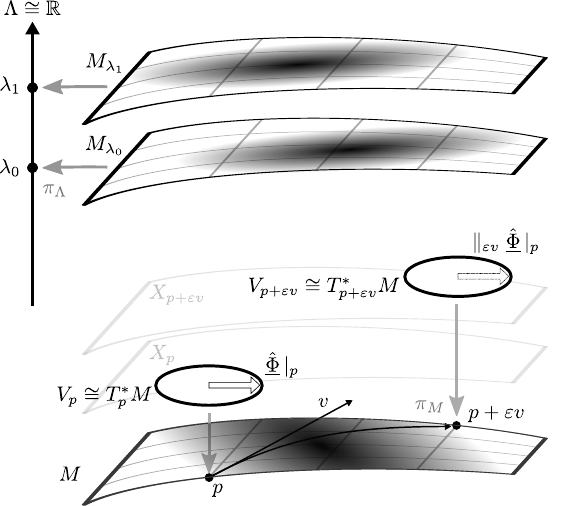}
\par\end{centering}
\caption{The Einstein nested bundle is an Einstein bundle congruence whose
collapsed bundle $M$ is also the base space of the cotangent bundle,
whose sections may be viewed as the matter fields of an embedded geometric
gauge theory whose fiber $X_{p}$ over every point $p\in M$ is isometric
to $M$. The matter sections of the Einstein bundle congruence define
a worldline congruence on $M$, whose rest density $\rho_{0}$ is
depicted by a gradient. The matter field collapses to a time-like
future-directed unit length cotangent vector field $\hat{\underline{\Phi}}$
on $M$ whose parallel transport is that of the spacetime connection
on $M$.}
\end{figure}

As an aside, we note that the matter field $\hat{\underline{\Phi}}$
at a point is a unit length cotangent vector, and may be contrasted
with $U$, which is not a tangent vector since it is metric-dependent;
$U$ may be instead viewed as the time-like vector in a tetrad. Thus
both $\hat{\underline{\Phi}}$ and $U$ are altered under a scaling
of the metric, since they are defined to be unit length, but only
$U$ is altered by a variation of the metric. It is also worth noting
that since it has no metric- and coordinate-independent form, and
since it cannot be varied freely due to being constrained to remain
divergenceless, we cannot consider the four-current to be a matter
field on spacetime.

\subsection{\label{subsec:GR-action}The action and equations of motion}

For the Einstein bundle, we may define the action to simply be that
of time-like dust with the addition of the spacetime scalar curvature:

\begin{equation}
\begin{aligned}S_{\mathrm{GTD}} & \equiv\int\left(-m\sqrt{-\left\langle J,J\right\rangle }+R\right)\mathrm{d}V\\
 & =\int\left(-m\sqrt{-\left\langle \mathfrak{J},\mathfrak{J}\right\rangle }+\sqrt{g}R\right)\mathrm{d}^{4}x.
\end{aligned}
\end{equation}
Variation of the worldlines yields the same EOM, constraining worldlines
to geodesics. Since there is no direct dependence upon the spacetime
connection, variation of parallel transport is equivalent to varying
the metric, which yields the usual EOM

\begin{equation}
\begin{aligned}\delta S_{\mathrm{GTD}}\left(g_{\mu\nu}\right) & =\int\left(\frac{1}{2}\frac{m}{\rho_{0}}J^{\mu}J^{\nu}-G^{\mu\nu}\right)\delta g_{\mu\nu}\sqrt{g}\mathrm{d}^{4}x\\
\Rightarrow G^{\mu\nu} & =\frac{1}{2}T_{\mathrm{TD}}^{\mu\nu},
\end{aligned}
\end{equation}
the Einstein field equations for time-like dust, where $G$ is the
Einstein tensor.

For the Einstein nested bundle, we define a different action

\begin{equation}
\begin{aligned}S_{\mathrm{G\Phi TD}} & \equiv\int\left(\rho_{0}\left(m\sqrt{-\left\langle \hat{\underline{\Phi}},\hat{\underline{\Phi}}\right\rangle }-m\right)+R\right)\mathrm{d}V,\end{aligned}
\end{equation}
and distinguish this theory by calling it \textbf{matter field gravitation}.
The variation of this action is less familiar, and we thus treat it
in a bit more detail. Variation of the matter field yields

\begin{equation}
\begin{aligned}\delta S_{\mathrm{G\Phi TD}}\left(\hat{\underline{\Phi}}\right) & =-m\rho_{0}\int\frac{\hat{\underline{\Phi}}^{\mu}}{\sqrt{-\left\langle \hat{\underline{\Phi}},\hat{\underline{\Phi}}\right\rangle }}\delta\hat{\underline{\Phi}}_{\mu}\mathrm{d}V=0\end{aligned}
\end{equation}
since $\delta\hat{\underline{\Phi}}$ is orthogonal to $\hat{\underline{\Phi}}$
for the variation of a unit length matter field. Hence we have no
EOM for the matter field. Variation of the underlying worldlines will
vary $\rho_{0}=\sqrt{-\left\langle J,J\right\rangle }$ per Section
\ref{subsec:Worldline-variation}, but we again obtain no EOM since 

\begin{equation}
\begin{aligned}\delta S_{\mathrm{G\Phi TD}}\left(x_{\alpha}^{\mu}\right) & =-\int\left(m\sqrt{-\left\langle \hat{\underline{\Phi}},\hat{\underline{\Phi}}\right\rangle }-m\right)\delta\rho_{0}\mathrm{d}V=0,\end{aligned}
\label{eq:GPhi-L-vanishes}
\end{equation}
due to the quantity in brackets vanishing. We are left with variation
of the metric, which yields
\begin{equation}
\begin{aligned}\delta S_{\mathrm{G\Phi TD}}\left(g_{\mu\nu}\right) & =m\rho_{0}\int\left(-\frac{1}{2}\hat{\underline{\Phi}}_{\mu}\hat{\underline{\Phi}_{\nu}}\delta g^{\mu\nu}-G^{\mu\nu}\delta g_{\mu\nu}\right)\sqrt{g}\mathrm{d}^{4}x\\
 & =\int\left(\frac{1}{2}m\rho_{0}\hat{\underline{\Phi}}^{\mu}\hat{\underline{\Phi}}^{\nu}-G^{\mu\nu}\right)\delta g_{\mu\nu}\mathrm{d}V,
\end{aligned}
\end{equation}
where the variations of $\rho_{0}$ and $\sqrt{g}$ for the first
term vanish since as in (\ref{eq:GPhi-L-vanishes}) they are multiplied
by zero, and we recall that $\delta g^{\mu\nu}=-g^{\mu\lambda}g^{\nu\sigma}\delta g_{\lambda\sigma}$.
This identifies the index-raised matter field as the unit vector in
the direction of the four-current, since 
\begin{equation}
\begin{aligned}\nabla_{\nu}\left(m\rho_{0}\hat{\underline{\Phi}}^{\mu}\hat{\underline{\Phi}}^{\nu}\right) & =m\hat{\underline{\Phi}}^{\mu}\nabla_{\nu}\left(\rho_{0}\hat{\underline{\Phi}}^{\nu}\right)+m\rho_{0}\hat{\underline{\Phi}}^{\nu}\nabla_{\nu}\hat{\underline{\Phi}}^{\mu}\\
 & =\nabla_{\nu}G^{\mu\nu}=0.
\end{aligned}
\end{equation}
In more detail, aligning our time-like coordinate with $\hat{\underline{\Phi}}^{\mu}$
at any point, $\hat{\underline{\Phi}}$ being of unit length means
that $\left(\nabla_{\nu}\hat{\underline{\Phi}}\right)^{0}=0$, and
consequently $\nabla_{\nu}\left(\rho_{0}\hat{\underline{\Phi}}^{\nu}\right)=0$,
which we therefore may identify with the four-current 
\begin{equation}
\begin{aligned}J & \equiv\rho_{0}\hat{\underline{\Phi}}^{\nu},\end{aligned}
\end{equation}
which due to the raised index has the proper metric dependence. This
in turn implies that in general the last term vanishes, which means
that the vector field $U^{\mu}\equiv\hat{\underline{\Phi}^{\mu}}$
satisfies $\nabla_{U}U=0$, i.e. the Einstein bundle worldlines follow
geodesics. Thus our metric EOM alone let us recover the worldline
EOM of the Einstein bundle and the divergenceless of the four-current.
In other words, the only SEM tensor of the form $fU^{\mu}U^{\nu}$
which can satisfy the Einstein field equations is $mU^{\nu}J^{\mu}$
where $\mathrm{div}J=0$ and $\nabla_{U}U=0$.

\subsection{\label{subsec:Units-GR}Units}

$R$ and $G$ are obtained from second derivatives of the metric tensor
components $g_{\mu\nu}$, which are dimensionless, so that 

\begin{equation}
\begin{aligned}\left[R\right]_{\mathrm{G}}=\left[G\right]_{\mathrm{G}} & =\mathsf{L}^{-2},\end{aligned}
\end{equation}
which remains so in SI units. The matter field is defined to be of
unit length, so $\hat{\underline{\Phi}}_{\mu}$ (and $\hat{\underline{\Phi}}^{\mu}$)
is dimensionless like $U^{\mu}$, and thus we have 
\begin{equation}
\begin{aligned}\left[\int\left(m\rho_{0}\sqrt{-g^{\mu\nu}\hat{\underline{\Phi}}_{\mu}\hat{\underline{\Phi}}_{\nu}}-m\rho_{0}+R\right)\mathrm{d}V\right]_{\mathrm{G}} & =\mathsf{L}^{2},\end{aligned}
\end{equation}
consistent with our other actions, while in SI units we use our conversion
rules to see that we again maintain units of energy-time

\begin{equation}
\begin{aligned}\left[\int\left(m\rho_{0}c^{2}\sqrt{-g^{\mu\nu}\hat{\underline{\Phi}}_{\mu}\hat{\underline{\Phi}}_{\nu}}-m\rho_{0}c^{2}+\frac{c^{4}}{16\pi k_{G}}R\right)\mathrm{d}V\right]_{\mathrm{SI}} & =\mathsf{M}\mathsf{L}^{2}\mathsf{T}^{-1}.\end{aligned}
\end{equation}

Now, in an orthonormal frame whose time-like component is aligned
with localized worldlines of unit particle number confined to an infinitesimal
sphere of radius $r$, the EOM are
\begin{equation}
G^{00}=\frac{1}{2}T_{\mathrm{TD}}^{00}=\frac{1}{2}m\rho_{0}.
\end{equation}
The other components of $T_{\mathrm{TD}}$ vanish, hence we also have
\begin{equation}
\begin{aligned}\sum G^{\mathsf{ii}} & =2\mathrm{Ric}(e_{0})-G^{00}=0\\
\Rightarrow\frac{1}{2}m\rho_{0} & =2\mathrm{Ric}(e_{0}),
\end{aligned}
\end{equation}
where $\mathrm{Ric}$ is the Ricci function. Recalling the geometrical
meaning of $\mathrm{Ric}$ from \cite{Marsh-book}, $\frac{1}{4}mr\rho_{0}$
is then the sum of the three accelerations $a$ of time-like geodesics
at the surface of the sphere towards each other, i.e. since the density
is one particle per sphere volume, we have
\begin{align}
\begin{aligned}\frac{1}{4}mr & =3a\frac{4}{3}\pi r^{3}\\
\Rightarrow a & =\frac{1}{16\pi}\frac{m}{r^{2}}.
\end{aligned}
\end{align}
This explains the conversion (\ref{eq:mass-conversion}), since it
results in
\begin{equation}
\begin{aligned}a_{\mathrm{SI}}c^{-2} & =\frac{1}{16\pi r^{2}}16\pi k_{G}c^{-2}m_{\mathrm{SI}}\\
\Rightarrow a_{\mathrm{SI}} & =\frac{k_{G}m_{\mathrm{SI}}}{r^{2}},
\end{aligned}
\label{eq:newton-force}
\end{equation}
Newton's law of universal gravitation from Section \ref{subsec:PM-action-and-variational-results}.
Linearizing general relativity maintains this relation at finite distances
(e.g. see \cite{Wald} pp. 76-78{]}).

\section{\label{sec:Matter-field-electromagnetism}Matter field electromagnetism }

In the previous paper \cite{Marsh-paper}, classical electromagnetism
was formulated as a geometric gauge theory we call \textbf{matter
field electromagnetism} (MFEM). Here we redefine this term to use
a new action and matter field which resolve some issues with the original
one, directly reduce to matter field gravitation, and result in consistent
geometric units. We assume familiarity with the geometric quantities
and results from \cite{Marsh-paper} without further comment.

\subsection{\label{subsec:The-Maxwell-nested-bundle}The Maxwell nested bundle }

The usual setting of electromagnetism is a $U(1)$ gauge theory absent
a matter field, which in \cite{Marsh-paper} we reformulated as a
$SO(2)$ gauge theory including a matter field. Under our present
definitions, this is a spacetime gauge theory which can be viewed
as a vector bundle $(\mathcal{X},M,\pi_{M},\mathbb{R}^{2})$, which
we call the \textbf{Maxwell bundle}.

Now, when considering units, it is clear that the matter field $\vec{\Phi}$
of \cite{Marsh-paper} cannot be viewed as a geometric quantity, since
it does not have units of length. One omission is that the dimensionful
constant $\hbar$ is not present, but inserting it (as we do below)
does not completely address this problem, which is compounded by the
odd dependence of the rest density on particle mass. To resolve these
issues, we define the base manifold $M$ of $\mathcal{X}$ to also
be the collapsed bundle of the Einstein bundle congruence $(\mathcal{E},\Lambda,\pi_{\Lambda},M)$,
as we did for the Einstein nested bundle. Also as in the Einstein
nested bundle, we define the matter field $\hat{\Phi}$ to be of unit
length, and preserve the rest density $\rho_{0}$ from the underlying
Einstein bundle congruence. We call this the \textbf{Maxwell nested
bundle}, and denote it $(\mathcal{X}_{\mathcal{E}},M,\pi_{M},\mathbb{R}^{2})$.
As we did in \cite{Marsh-paper}, in order to map to other treatments
and use the gauge potential $A_{\mu}$, we will utilize complex notation
in parallel with two dimensional vector notation.
\begin{figure}[H]
\noindent \begin{centering}
\includegraphics{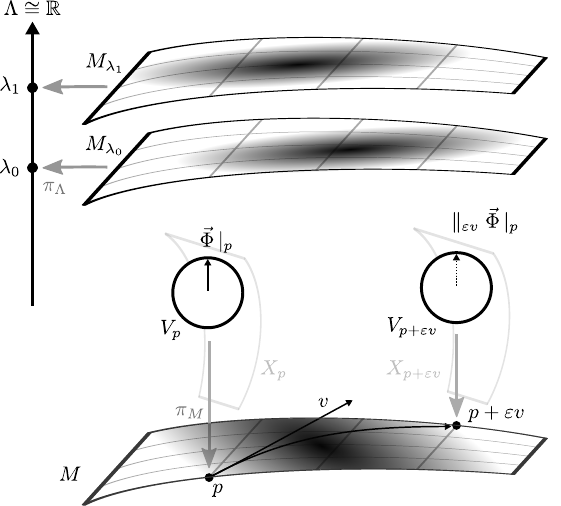}
\par\end{centering}
\caption{The Maxwell nested bundle is an Einstein bundle congruence $(\mathcal{E},\Lambda,\pi_{\Lambda},M)$
whose collapsed bundle $M$ is also the base space of the Maxwell
bundle, a spacetime geometric gauge theory whose two-dimensional fiber
$X_{p}$ allows us to view it as a vector bundle $(\mathcal{X},M,\pi_{M},\mathbb{R}^{2})$. }
\end{figure}

One might wonder whether it makes sense to continue the analogy with
matter field gravitation, and form an embedded version of the Maxwell
nested bundle, wherein the internal space is a space-like cotangent
space in spacetime. The intention is to show in a subsequent paper
that this describes the geometric setting of classical Dirac theory. 

\subsection{\label{subsec:EM-action}The action and equations of motion}

In order to map to complex notation, we define 
\begin{align}
\begin{aligned}\vec{\Phi} & \equiv\sqrt{\rho_{0}}\hat{\Phi}\\
 & \leftrightarrow\Phi\in\mathbb{C},
\end{aligned}
\end{align}
which we emphasize cannot be considered a geometric matter field,
since $\left[\Phi^{a}\right]_{\mathrm{G}}=\mathsf{L}^{-\nicefrac{3}{2}}$.
We define the Lagrangian for MFEM as:
\begin{align}
\begin{aligned}L_{\mathrm{EM}} & \equiv\rho_{0}\left(\hbar\sqrt{-\left\langle \mathrm{D}_{\mu}\hat{\Phi},\mathrm{D}_{\mu}\hat{\Phi}\right\rangle }-m\right)+\frac{\hbar}{16\pi\alpha}\left\langle R_{A},R_{A}\right\rangle +R\\
 & =\hbar\left\langle \vec{\Phi},\vec{\Phi}\right\rangle \sqrt{-\left\langle \mathrm{D}_{\mu}^{\varangle}\vec{\Phi},\mathrm{D}_{\mu}^{\varangle}\vec{\Phi}\right\rangle }-m\left\langle \vec{\Phi},\vec{\Phi}\right\rangle +\frac{\hbar}{16\pi\alpha}\left\langle R_{A},R_{A}\right\rangle +R\\
 & =\frac{\hbar}{4}\sqrt{\Phi^{*}\Phi\left(\Phi^{*}\overleftrightarrow{\mathrm{D}}_{\mu}\Phi\right)\left(\Phi^{*}\overleftrightarrow{\mathrm{D}}^{\mu}\Phi\right)}-m\Phi^{*}\Phi-\frac{1}{4}\varepsilon_{0}F_{\mu\nu}F^{\mu\nu}+R,
\end{aligned}
\end{align}
where
\begin{equation}
\begin{aligned}\Phi^{*}\overleftrightarrow{\mathrm{D}}_{\mu}\Phi & \equiv\Phi^{*}\mathrm{D}_{\mu}\Phi-\mathrm{D}_{\mu}\Phi^{*}\Phi,\end{aligned}
\end{equation}
and we note that since the infinitesimal displacement of a unit length
vector is equal to the infinitesimal rotation in radians, we have
\begin{equation}
\mathrm{D}_{\mu}^{\varangle}\vec{\Phi}=\mathrm{D}_{\mu}\hat{\Phi}.
\end{equation}
The dimensionless $\alpha$ is the fine structure constant, and 
\begin{align}
\begin{aligned}\varepsilon_{0} & \equiv\frac{e^{2}}{4\pi\hbar\alpha}\end{aligned}
\end{align}
is the permittivity of free space, which is defined in terms of the
electron charge $e=q$. Note that in the first two versions of the
Lagrangian, $e$ cancels with $q$; as we will verify, this is consequently
true for all the EOM, and hence the value and units of both $e$ and
$q$ are arbitrary. Going forward we define both to be unity and dimensionless,
but note that it is common to take their geometrized units to be length
in order to make $\varepsilon_{0}$ and therefore $k_{e}$ dimensionless. 

The canonical momentum is 
\begin{align}
\begin{aligned}p_{\hat{\Phi}a}^{\mu}=\frac{\partial L_{\mathrm{EM}}}{\partial\left(\mathrm{D}_{\mu}\hat{\Phi}^{a}\right)} & =-\rho_{0}\hbar\frac{\mathrm{D}^{\mu}\hat{\Phi}_{a}}{\sqrt{-\left\langle \mathrm{D}_{\mu}\hat{\Phi},\mathrm{D}_{\mu}\hat{\Phi}\right\rangle }}\\
\Rightarrow\hat{\Phi}^{a}p_{\hat{\Phi}a}^{\mu} & =0,
\end{aligned}
\end{align}
since the change in the unit length $\hat{\Phi}$ must be orthogonal
to it. Thus by Euler's homogeneous function theorem, on-shell we have
\begin{equation}
\hbar\sqrt{-\left\langle \mathrm{D}_{\mu}\hat{\Phi},\mathrm{D}_{\mu}\hat{\Phi}\right\rangle }=m,
\end{equation}
i.e. the Euler-Lagrange equation is
\begin{equation}
\hbar\mathrm{D}_{U}\hat{\Phi}=\pm m,
\end{equation}
where $U$ is defined to be the unit length time-like future-directed
four-vector in the direction of extremal counterclockwise matter field
angular velocity. We can also verify this result via direct calculations
as in \cite{Marsh-paper}, whether in terms of $\hat{\Phi}$, $\vec{\Phi}$,
or $\Phi$. Note that $\hbar$ then has a geometrical interpretation
as the mass per unit of matter field angular velocity, while the\textbf{
}Compton wavelength $\lambda=2\pi\hbar/m$ is the time-like distance
in which the matter field rotates by one revolution.

The vanishing of the first term in the Lagrangian on-shell means that
as in matter field gravitation, variation of $\rho_{0}$ via variation
of the Einstein bundle worldlines results in no EOM. Recalling from
\cite{Marsh-paper} that 
\begin{align}
\begin{aligned}\mathrm{D}_{\mu}^{\varangle}\Phi & =\frac{-i\left(\Phi^{*}\partial_{\mu}\Phi-\Phi\partial_{\mu}\Phi^{*}\right)}{2\left|\Phi\right|^{2}}-\qbar A_{\mu}\\
\Rightarrow\frac{\partial\left(\mathrm{D}_{\mu}^{\varangle}\Phi\right)}{\partial A_{\mu}} & =-\qbar,
\end{aligned}
\end{align}
we have
\begin{equation}
\begin{aligned}\frac{\partial L_{\mathrm{EM}}}{\partial A_{\nu}} & =\hbar\rho_{0}\frac{\qbar\mathrm{D}^{\varangle\nu}\Phi}{\sqrt{-\left\langle \mathrm{D}_{\mu}^{\varangle}\Phi,\mathrm{D}_{\mu}^{\varangle}\Phi\right\rangle }}\\
 & =\mathrm{sign}\left(\mathrm{D}^{\varangle\nu}\Phi\right)q\rho_{0}U^{\nu}.
\end{aligned}
\end{equation}
Note that the sign factor is necessary to ensure that $U$ is future-directed.
We define 
\begin{equation}
\begin{aligned}Q & \equiv\mathrm{sign}\left(\mathrm{D}^{\varangle\nu}\Phi\right)q,\end{aligned}
\end{equation}
which is what is usually referred to as the electric charge per particle;
it is positive when the angular velocity of the matter field is in
the clockwise direction, since then
\begin{equation}
\begin{aligned}\mathrm{D}_{U}^{\varangle}\Phi & =U^{\nu}\mathrm{D}_{\nu}^{\varangle}\Phi<0\\
\Rightarrow\mathrm{D}_{\nu}^{\varangle}\Phi & <0\\
\Rightarrow\mathrm{D}^{\varangle\nu}\Phi & >0.
\end{aligned}
\end{equation}
On-shell, the sign of the matter field EOM thus corresponds to the
sign of $Q$:
\begin{equation}
\begin{aligned}\hbar\mathrm{D}_{U}^{\varangle}\Phi & =-m & \Rightarrow Q & =+1 & \textrm{} & \textrm{clockwise}\\
\hbar\mathrm{D}_{U}^{\varangle}\Phi & =+m & \Rightarrow Q & =-1 & \textrm{} & \textrm{counter-clockwise}
\end{aligned}
\end{equation}
Referring to \cite{Marsh-paper} the Euler-Lagrange equation is then
\begin{equation}
\begin{aligned}Q\rho_{0}U^{\nu} & =\varepsilon_{0}\nabla_{\mu}F^{\nu\mu}\\
\Rightarrow J^{\nu} & =\rho_{0}U^{\nu},
\end{aligned}
\end{equation}
and the four-current has the correct metric dependence since $U^{\nu}$
is an index-raised 1-form. Again, these results can be verified in
terms of $\hat{\Phi}$ or $\vec{\Phi}$. 

Note that in pure geometric terms we have
\begin{equation}
\begin{aligned}\mathrm{sign}\left(\mathrm{D}^{\varangle\nu}\Phi\right)\left(\frac{q}{e}\right)^{2}\rho_{0}U^{\nu} & =\frac{1}{4\pi\alpha}\nabla_{\mu}\check{R}_{A}^{\nu\mu},\end{aligned}
\label{eq:geom-maxwell}
\end{equation}
so that the curvature responds to the square of the coupling constant
measured in units of electron charge, with the sign dependent upon
the direction of rotation. As promised, the value and units of $q=e$
are thus arbitrary. If we approximate spacetime as flat, and at a
given time $\rho_{0}$ integrates to unity over a space-like sphere
of radius $r$ and vanishes outside the sphere, then in inertial coordinates
whose time axis is aligned with $U$, we can integrate (\ref{eq:geom-maxwell})
over the sphere to yield 
\begin{equation}
\begin{aligned}\frac{\alpha}{r^{2}} & \overset{\mathbb{M}}{=}\check{R}_{A}^{\mathsf{j}0},\end{aligned}
\label{eq:}
\end{equation}
where we use the divergence theorem and the symmetry of the configuration.
This is a geometric form of Coulomb's law, with the fine structure
constant acting as the ``geometric charge.'' 

Again referring to \cite{Marsh-paper}, variation of the metric yields
\begin{equation}
\begin{aligned}\delta S_{\mathrm{EM}}\left(g_{\mu\nu}\right) & =\int\left(\frac{\rho_{0}\hbar\mathrm{D}^{\varangle\mu}\vec{\Phi}\mathrm{D}^{\varangle\nu}\vec{\Phi}}{2\sqrt{-\left\langle \mathrm{D}_{\mu}^{\varangle}\vec{\Phi},\mathrm{D}_{\mu}^{\varangle}\vec{\Phi}\right\rangle }}+\frac{1}{2}T_{\mathrm{EM}}^{\mu\nu}-G^{\mu\nu}\right)\delta g_{\mu\nu}\sqrt{g}\mathrm{d}^{4}x\\
 & \overset{\Phi_{\mathrm{EL}}}{=}\frac{1}{2}\int\left(m\rho_{0}U^{\mu}U^{\nu}+T_{\mathrm{EM}}^{\mu\nu}-G^{\mu\nu}\right)\delta g_{\mu\nu}\sqrt{g}\mathrm{d}^{4}x,
\end{aligned}
\end{equation}
where again the variations of $\rho_{0}$ and $\sqrt{g}$ for the
first term vanish since as in (\ref{eq:GPhi-L-vanishes}) they are
multiplied by zero on-shell; thus as in \cite{Marsh-paper} we recover
the correct Hilbert SEM tensor and the Lorentz force law. 

Note that the QED approximation in \cite{Marsh-paper} remains valid,
in that it results in EOM which continue to match those of the improved
MFEM. Also note that our new Lagrangian is obtained from $L_{\mathrm{DIRAC-\Phi}}$
by taking absolute values instead of squares. Finally, constructing
an electrically neutral four-current via two matter fields 
\begin{equation}
\hbar\mathrm{D}_{U}^{\varangle}\hat{\Phi}_{+}=-m,\qquad\hbar\mathrm{D}_{U}^{\varangle}\hat{\Phi}_{-}=m
\end{equation}
with parallel four-currents and each with rest density $\frac{1}{2}\rho_{0}$,
results in a total $QJ$ and thus $F$ which vanishes, and an on-shell
Lagrangian
\begin{equation}
L_{\mathrm{EM\pm}}\overset{\Phi_{\mathrm{EL}}}{=}\rho_{0}\left(m\sqrt{-\left\langle U_{\mu},U_{\mu}\right\rangle }-m\right)+R
\end{equation}
which is identical to that of matter field gravitation.

\subsection{\label{subsec:Units-EM}Units}

As we saw in Section \ref{subsec:EM-action}, the reduced Planck constant
$\hbar$ is the mass per unit of matter field angular velocity. Since
relativistic angular velocity is radians per length, we have 
\begin{equation}
\begin{aligned}\left[\mathrm{D}_{\mu}^{\varangle}\vec{\Phi}\right]_{\mathrm{G}} & =\mathsf{L}^{-1}=\left[\mathrm{D}_{\mu}\hat{\Phi}\right]_{\mathrm{G}}\\
\Rightarrow\left[\hbar\right]_{\mathrm{G}} & =\mathsf{L}^{2}\\
\Rightarrow\left[\varepsilon_{0}\right]_{\mathrm{G}} & =\mathsf{L}^{-2}.
\end{aligned}
\end{equation}
As mentioned in Section \ref{subsec:The-Maxwell-nested-bundle}, $\hbar$
is thus geometrically dimensionful, and setting it to unity means
defining a length scale. 

Now, the gauge potential, as part of the gauge covariant derivative,
must have
\begin{equation}
\begin{aligned}\left[-i\frac{q}{\hbar}A_{\mu}\right]_{\mathrm{G}} & =\mathsf{L}^{-1}\\
\Rightarrow\left[A_{\mu}\right]_{\mathrm{G}} & =\mathsf{L}\\
\Rightarrow\left[F_{\mu\nu}\right]_{\mathrm{G}} & =\mathsf{1},
\end{aligned}
\end{equation}
since $i$ is a unit basis of $u(1)$ and is therefore dimensionless.
Hence, 

\begin{equation}
\begin{aligned}\left[\int\left(\rho_{0}\left(\hbar\sqrt{-\left\langle \mathrm{D}_{\mu}\hat{\Phi},\mathrm{D}_{\mu}\hat{\Phi}\right\rangle }-m\right)-\frac{1}{4}\varepsilon_{0}F_{\mu\nu}F^{\mu\nu}\right)\mathrm{d}V\right]_{\mathrm{G}} & =\mathsf{L}^{2},\end{aligned}
\end{equation}
consistent with our other actions. 

In SI units, the fine structure constant remains dimensionless, but
$q$ (and therefore $Q$) has its own base physical dimension of charge
$\mathsf{Q}$ (although the base dimension sometimes used is current
$\mathsf{I}=\mathsf{QT}$). Hence, since the first term in the Lagrangian
acquires a $c$ from the time-like derivative
\[
\begin{aligned}\left\langle \mathrm{D}_{\mu}\hat{\Phi},\mathrm{D}_{\mu}\hat{\Phi}\right\rangle _{\mathrm{G}} & \rightarrow c^{2}\left\langle \mathrm{D}_{\mu}\hat{\Phi},\mathrm{D}_{\mu}\hat{\Phi}\right\rangle _{\mathrm{SI}},\end{aligned}
\]
while the second gains a $c$ per our conversion rules, to keep them
consistent we have
\begin{equation}
\begin{aligned}\left[\hbar\right]_{\mathrm{SI}} & =\mathsf{L}\left[mc\right]_{\mathrm{SI}}=\mathsf{M}\mathsf{L}^{2}\mathsf{T}^{-1}\\
\Rightarrow\left[A_{\mu}\right]_{\mathrm{SI}} & =\mathsf{Q}^{-1}\mathsf{M}\mathsf{L}\mathsf{T}^{-1}\\
\Rightarrow\left[F_{\mu\nu}\right]_{\mathrm{SI}} & =\mathsf{Q}^{-1}\mathsf{M}\mathsf{T}^{-1}.
\end{aligned}
\end{equation}
The constant $\varepsilon_{0}$ also gains a factor of $c$ in the
denominator to match the units of the Coulomb constant from (\ref{eq:coulomb-const-units})
\begin{equation}
\begin{aligned}\left[\varepsilon_{0}=\frac{1}{4\pi k_{e}}\right]_{\mathrm{SI}} & =\mathsf{Q}^{2}\mathsf{M}^{-1}\mathsf{L}^{-3}\mathsf{T}^{2}\\
\Rightarrow\left(\varepsilon_{0}\right)_{\mathrm{SI}} & \equiv\frac{e^{2}}{4\pi\hbar c\alpha},
\end{aligned}
\end{equation}
resulting in an action with consistent units of energy-time

\begin{equation}
\begin{aligned}\left[\int\left(\rho_{0}\left(\hbar c\sqrt{-\left\langle \mathrm{D}_{\mu}\hat{\Phi},\mathrm{D}_{\mu}\hat{\Phi}\right\rangle }-mc^{2}\right)-\frac{1}{4}\varepsilon_{0}c^{2}F_{\mu\nu}F^{\mu\nu}\right)\mathrm{d}V\right]_{\mathrm{SI}} & =\mathsf{M}\mathsf{L}^{2}\mathsf{T}^{-1}.\end{aligned}
\end{equation}

\section{\label{sec:Summary}Summary}

Having constructed geometric gauge theories corresponding to classical
theories of physics, from particle mechanics to electromagnetism,
we now may summarize the limits under which each theory yields the
next in the reverse sequence.{\footnotesize{}}
\begin{table}[H]
{\footnotesize{}}%
\begin{tabular*}{1\columnwidth}{@{\extracolsep{\fill}}>{\raggedright}p{0.48\columnwidth}>{\raggedright}p{0.27\columnwidth}>{\raggedright}p{0.25\columnwidth}}
\toprule 
{\footnotesize{}Action} & {\footnotesize{}Equations of motion} & {\footnotesize{}Bundles}\tabularnewline
\midrule
\midrule 
{\footnotesize{}Electromagnetism} &  & {\footnotesize{}Maxwell nested bundle}\tabularnewline
\addlinespace[0.4cm]
{\footnotesize{}$\delta\int\left(\rho_{0}(\hbar\left\Vert \mathrm{D}_{\mu}\hat{\Phi}\right\Vert -m)+R-\frac{\hbar}{16\pi\alpha}\left\Vert R_{A}\right\Vert ^{2}\right)\mathrm{d}V$\qquad{}$=\delta\int\left(\left\Vert \Phi\right\Vert (\frac{\hbar}{4}\left\Vert \Phi^{*}\overleftrightarrow{\mathrm{D}}\Phi\right\Vert -m)+R-\frac{1}{2}\varepsilon_{0}\left\Vert F\right\Vert ^{2}\right)\mathrm{d}V$} & {\footnotesize{}$\hbar\mathrm{D}_{U}\hat{\Phi}=\pm m$, $Q\rho_{0}U^{\nu}=\varepsilon_{0}\nabla_{\mu}F^{\nu\mu}$,
$m\rho_{0}U^{\mu}U^{\nu}+T_{\mathrm{EM}}^{\mu\nu}=G^{\mu\nu}$} & {\footnotesize{}$(\mathcal{X}_{\mathcal{E}},M,\pi_{M},\mathbb{R}^{2})$,
$(\mathcal{E},\Lambda,\pi_{\Lambda},M)$}\tabularnewline
\addlinespace[0.4cm]
\midrule 
\multicolumn{3}{c}{{\footnotesize{}$\hat{\Phi}\rightarrow\frac{1}{2}\left(\hat{\Phi}_{+}+\hat{\Phi}_{-}\right)\Rightarrow\hbar\mathrm{D}_{\mu}\hat{\Phi}_{\pm}\rightarrow\pm m\hat{\underline{\Phi}}_{\mu},\ F\rightarrow0,\ T_{\mathrm{EM}}\rightarrow0$}}\tabularnewline
{\footnotesize{}General relativity} & {\footnotesize{}$\Downarrow$} & {\footnotesize{}Einstein nested bundle}\tabularnewline
\addlinespace[0.4cm]
{\footnotesize{}$\delta\int\left(\rho_{0}(m\left\Vert \hat{\underline{\Phi}}\right\Vert -m)+R\right)\mathrm{d}V$\qquad{}\qquad{}$=\delta\int\left(-m\left\Vert J\right\Vert +R\right)\mathrm{d}V$} & {\footnotesize{}$m\rho_{0}\hat{\underline{\Phi}}^{\mu}\hat{\underline{\Phi}}^{\nu}=G^{\mu\nu}$
$\Rightarrow U^{\mu}\equiv\hat{\underline{\Phi}}^{\mu},\ \nabla_{U}U=0$} & {\footnotesize{}$(T^{*}M_{\mathcal{E}},M,\pi_{M},\mathbb{R}^{3,1}),$
$(\mathcal{E},\Lambda,\pi_{\Lambda},M)$}\tabularnewline
\addlinespace[0.4cm]
\midrule 
\multicolumn{3}{c}{{\footnotesize{}$R\rightarrow0,\ \hat{\underline{\Phi}}\rightarrow U\equiv\mathrm{D}_{\tau}\sigma\Rightarrow\rho_{0}\hat{\underline{\Phi}}\rightarrow J$}}\tabularnewline
{\footnotesize{}Relativistic continua} & {\footnotesize{}$\Downarrow$} & {\footnotesize{}Minkowski bundle congruence}\tabularnewline
\addlinespace[0.4cm]
{\footnotesize{}$\delta\int\frac{1}{2}\rho_{0}\left(-m\left\Vert \mathrm{D}_{\tau}\sigma\right\Vert -m\right)\mathrm{d}V$\qquad{}\qquad{}\qquad{}\qquad{}\qquad{}$=-\delta\int m\left\Vert J\right\Vert \mathrm{d}V$} & {\footnotesize{}$\nabla_{U}U=0$} & {\footnotesize{}$(\mathcal{M},\Lambda,\pi,\mathbb{M})$}\tabularnewline
\addlinespace[0.4cm]
\midrule 
\multicolumn{3}{c}{{\footnotesize{}$\begin{aligned}L\mathrm{d}V & \rightarrow\frac{L}{\rho_{0}}\mathrm{d}\tau\end{aligned}
$}}\tabularnewline
{\footnotesize{}Relativistic particles} & {\footnotesize{}$\Downarrow$} & {\footnotesize{}Minkowski bundle}\tabularnewline
\addlinespace[0.4cm]
{\footnotesize{}$\delta\int\frac{1}{2}\left(-m\left\Vert \mathrm{D}_{\tau}\sigma\right\Vert -m\right)\mathrm{d}\tau$
\qquad{}\qquad{}\qquad{}\qquad{}$=-\delta\int m\left\Vert P\right\Vert \mathrm{d}\lambda$} & {\footnotesize{}$\partial_{\lambda}P^{\mu}=0$} & {\footnotesize{}$(\mathcal{M},\Lambda,\pi,\mathbb{M})$}\tabularnewline
\addlinespace[0.4cm]
\midrule 
\multicolumn{3}{c}{{\footnotesize{}$\begin{aligned}L\left(\lambda\right)\mathrm{d}\lambda & \rightarrow cL\left(t\right)\mathrm{d}t,\ \left\Vert \mathrm{D}_{t}\sigma\right\Vert =\sqrt{c^{2}-v^{2}}\rightarrow\left(c-\frac{1}{2}\frac{v^{2}}{c}\right)\end{aligned}
$}}\tabularnewline
\addlinespace
{\footnotesize{}Particle mechanics} & {\footnotesize{}$\Downarrow$} & {\footnotesize{}Galilean nested bundle}\tabularnewline
\addlinespace[0.4cm]
{\footnotesize{}$\delta\int\left(\frac{1}{2}m\left\Vert \mathrm{D}_{t}\sigma\right\Vert ^{2}-m\right)\mathrm{d}t$
\qquad{}\qquad{}\qquad{}\qquad{}\qquad{}$=\delta\int\frac{1}{2}mv^{2}\mathrm{d}t$} & {\footnotesize{}$ma=0$} & {\footnotesize{}$(E_{T}^{\mathcal{G}},T,\pi_{\mathcal{G}},\mathcal{G})$,\qquad{}
$(\mathcal{G},T,\pi_{S},S)$}\tabularnewline\addlinespace[0.2cm]
\bottomrule
\addlinespace[0.4cm]
\end{tabular*}{\footnotesize\par}

{\footnotesize{}\caption{Two EM matter fields with the same rest density but opposite rotations
result in zero net charge four-current and thus zero electromagnetic
field, with the EOM enabling the definition of a unit 1-form field
to yield GR; taking the limit of spacetime as flat and defining the
index-raised 1-form field as the direction of the four-current of
the worldline congruence yields relativistic continua; taking the
limit of a single particle number along a single worldline yields
relativistic particles; and dropping the constant Lagrangian term,
changing the integration variable to a time coordinate, and taking
the limit of $v<<c$ in an expansion to order $v^{2}$ yields particle
mechanics.}
}{\footnotesize\par}
\end{table}
{\footnotesize\par}

\end{document}